\documentclass{article}

\usepackage[a4paper]{geometry}

\usepackage{graphicx}
\usepackage{adjustbox}
\usepackage[colorlinks=true, allcolors=blue]{hyperref}
\usepackage{amsmath}
\usepackage{kmath}
\usepackage{mathtools} 
\usepackage{txfonts}
\usepackage{amssymb}
\usepackage[version=4]{mhchem}
\usepackage{siunitx}
\DeclareSIUnit\angstrom{\text {Å}}

\usepackage{bbding}
\usepackage{setspace}
\usepackage[utf8]{inputenc} 
\usepackage{placeins} 
\usepackage{pifont} 
\usepackage{xcolor} 
\definecolor{orange}{RGB}{253,135,2}
\definecolor{bgd}{RGB}{128,0,32}
\definecolor{pp}{RGB}{190,0,224}
\usepackage[normalem]{ulem} 
\usepackage{authblk} 
\usepackage{verbatim} 
\usepackage[numbers,sort&compress]{natbib}
\usepackage{indentfirst} 
\usepackage{tabularray} 
\usepackage{textgreek}
\usepackage{rotating}
\usepackage{multirow}
\usepackage{colortbl}
\usepackage{kotex}

\newcommand{\super}[1]{\textsuperscript{#1}}
\newcommand{\sub}[1]{\textsubscript{#1}}
\newcommand{\mul}{$\times$}
\newcommand{\plm}{$\pm$}

\newcommand{\um}{$\mu$m}

\newcommand{\ca}{$\sim$}
\newcommand{\degC}{°C}
\newcommand{\invcm}{cm\textsuperscript{-1}}
\newcommand{\excite}{\textlambda\textsubscript{ex}}
\newcommand{\sptwo}{\textit{sp}\textsuperscript{2}}
\newcommand{\TTC}{T\textsubscript{TC}}
\newcommand{\TPy}{T\textsubscript{Py}}
\newcommand{\pie}{$\pi$}

\DeclareSIUnit{\rydberg}{Ry}
\DeclareSIUnit{\bohr}{Bohr}

\title{Ultra-high vacuum Raman platform for in situ characterization of graphene nanoribbons}

\author[1,2]{Jeong Ha Hwang}
\author[1]{Amogh Kinikar\thanks{Current affiliation: Karlsruhe Institute of Technology, 76131 Karlsruhe, Germany}}
\author[1]{Lukas Rotach}
\author[1]{Andres Ortega-Guerrero}
\author[1]{Carlo A. Pignedoli}
\author[3]{Klaus M\"ullen}
\author[4]{Thorsten G. Englmann}
\author[4,5]{Xinliang Feng}
\author[1]{Pascal Ruffieux}
\author[1,6]{Roman Fasel}
\author[2,7,8]{Mickael L. Perrin}
\author[1]{Gabriela Borin Barin\thanks{Corresponding author: Gabriela Borin Barin, gabriela.borin-barin@empa.ch}}

\affil[1]{nanotech@surfaces Laboratory, Empa - Swiss Federal Laboratories for Materials Science and Technology, 8600 D\"ubendorf, Switzerland}
\affil[2]{Department of Information Technology and Electrical Engineering, ETH Z\"urich - Swiss Federal Institute of Technology Z\"urich, 8092 Z\"urich, Switzerland}
\affil[3]{Max Planck Institute for Polymer Research, 55128 Mainz, Germany}
\affil[4]{Faculty of Chemistry and Food Chemistry \& Center for Advanced Electronics Dresden, TUD - Dresden University of Technology, 01062 Dresden, Germany}
\affil[5]{Max Planck Institute of Microstructure Physics, 06120 Halle, Germany}
\affil[6]{Department of Chemistry, Biochemistry, and Pharmaceutical Sciences, University of Bern, 3012 Bern, Switzerland}
\affil[7]{Transport at Nanoscale Interfaces Laboratory, Empa - Swiss Federal Laboratories for Materials Science and Technology, 8600 D\"ubendorf, Switzerland}
\affil[8]{Quantum Center, ETH Z\"urich - Swiss Federal Institute of Technology Z\"urich, 8093 Z\"urich, Switzerland}

\begin{document}

\maketitle
\abstract{
Atomically precise graphene nanoribbons (GNRs) exhibit tunable electronic and magnetic properties governed by edge topology and finite-size effects, which make them versatile platforms for next-generation electronic and spintronic applications. 
However, the unpaired \pie-electrons responsible for their magnetic properties simultaneously make them highly susceptible to chemical degradation under ambient conditions. 
This intrinsic reactivity poses a central experimental challenge: accessing vibrational and electronic signatures of air-sensitive GNRs during synthesis and under controlled environments without breaking vacuum. 
Once the material has been exposed to air, standard characterization techniques would probe oxidized or chemically modified species rather than the pristine form. 
Here, we overcome this limitation by developing a home-built ultra-high vacuum (UHV) Raman platform designed to preserve sample integrity by preventing air exposure and to enable in situ investigation of material properties. 
The portable Raman vacuum suitcase (RVS) integrates temperature control and precise gas dosing, allowing direct monitoring of growth kinetics, lattice dynamics, and reactive-edge responses under well-defined thermal and chemical environments. 
Using this platform, we monitor the on-surface synthesis of 7- and 9-atom-wide armchair GNRs (7- and 9-AGNRs), quantify the evolution of 7-AGNR Raman modes over a wide temperature range (162--748 K), and resolve chemical changes upon controlled \ce{O2} exposure that are consistent with oxidation at the reactive zigzag sites. 
These results establish UHV Raman spectroscopy with the RVS as a route to accessing the intrinsic vibrational signatures of low-dimensional quantum materials under controlled environments. 
}

\section{Introduction}
One- and two-dimensional (1D and 2D) quantum materials are a rapidly expanding class of systems in which quantum confinement reshapes electronic, vibrational, and magnetic properties\cite{Geim2007rise,Houtsma2021Atomically,Guo2021One,Song2021properties,Cai2010Atomically}. 
Such confinement effects are particularly pronounced when moving from bulk crystals to 2D systems such as graphene, black phosphorus, and transition metal dichalcogenides\cite{Burch2018Magnetism,Zhang2021LayerDependent,Mak2010Atomically}. 
Graphene, in particular, exhibits high thermal and electrical conductivities\cite{Balandin2008Superior, Bolotin2008Ultrahigh, Castro2009electronic}. Further reduction of dimensionality to 1D derivatives such as carbon nanotubes (CNTs) or graphene nanoribbons (GNRs) introduces an additional layer of tunability. 
In CNTs, the electronic structure is governed by the tube diameter and chirality, both defined by the rolling direction of the graphene sheet\cite{Iijima1991Helical,Iijima1993Single,Charlier2007Electronic, Saito1992Electronic}. 
In GNRs, the ribbon width, edge structure, termini, and lattice symmetry give rise to electronic and magnetic properties absent in graphene\cite{Son2006Energy,Son2006Half,Houtsma2021Atomically, Ruffieux2016On}. 
Exploiting this tunability, however, requires synthetic control at the atomic level since even minor structural variations can lead to pronounced changes in the material properties. 

On-surface synthesis (OSS) enables precise engineering of GNR width and edge topology\cite{Groning2018Engineering,Rizzo2018Topological}, providing a direct route to tailor their electronic\cite{Houtsma2021Atomically}, magnetic\cite{Blackwell2021Spin}, and optical properties\cite{Nascimento2025Optical}. 
Many of these properties originate from unpaired \pie-electrons, such as those at zigzag edge segments. 
These electrons give rise to spin-polarized edge states with strong potential for spintronic applications\cite{Wang2021Graphene,Zhang2026Bottomup}. 
The open-shell configuration of these states, however, renders such GNRs highly reactive toward ambient species, limiting their stability upon exposure to air\cite{Lawrence2022Circumventing}. 
Consequently, the very features that make them promising for spintronics also limit their characterization to ultra-high vacuum (UHV) environments\cite{Lawrence2022Circumventing,Berdonces2021Chemical}.

Outside UHV, GNRs are susceptible to uncontrolled adsorption, functionalization, and oxidation. 
Hydroxyl, ketone, and ether groups, as well as additional hydrogen atoms, can be randomly introduced, fundamentally altering their chemical structure\cite{Berdonces2021Chemical}. 
Measurements performed under ambient conditions or after exposure to air therefore probe a chemically altered system and yield vibrational, electronic, and magnetic signatures that no longer reflect the pristine material\cite{Sykora2010Effect,Berdonces2021Chemical}. 
To date, this high reactivity has limited device integration primarily to armchair GNRs (AGNRs). Their closed-shell electronic structure\cite{Son2006Energy} has so far provided sufficient stability against air, acid fumes\cite{Fairbrother2017High,Borin2019Surface}, and polymers\cite{Llinas2017Short,Zhang2023Contacting} encountered during transfer from metallic growth substrates to technologically relevant dielectric platforms\cite{El2020Controlled,Llinas2017Short}.

Overcoming the challenges of integrating reactive GNRs into devices requires new workflows that preserve their structural integrity throughout fabrication. 
This, in turn, demands characterization tools capable of assessing the chemical structure of these carbon nanomaterials after each step of the integration process. 
Raman spectroscopy is a non-destructive technique that provides chemically specific information through inelastic light scattering, resolving lattice dynamics, edge-related vibrational modes, and chemical bonding in low-dimensional materials\cite{Ferrari2013Raman, Gillen2009Vibrational, Verzhbitskiy2016Raman}. 
In the context of GNRs, it has proven particularly powerful for monitoring structural quality and integrity both on the growth substrate and after device integration\cite{Overbeck2019Universal,Borin2019Surface,Zhang2026Bottomup}.

Extending this diagnostic capability to GNRs hosting spin-polarized edge states or topological quantum states is crucial for bringing these structures toward device integration. 
However, the lack of a broadly adaptable, environmentally isolated spectroscopic platform has so far limited access to the pristine vibrational fingerprints of these systems. 
Without reliable access to their intrinsic modes, neither structural evolution nor integrity can be tracked across the substrate transfer steps required for device integration. 
Establishing a dedicated Raman methodology under UHV conditions is therefore essential to overcome this challenge and provide the metrology needed to unlock the device potential of reactive GNRs.

Indeed, Raman spectroscopy has already been used to track the vibrational signatures of GNRs as a function of temperature\cite{Guo2022Phonon} and recent UHV studies have characterized GNRs at room and cryogenic temperatures\cite{Gruneis2018Ultrahigh,Senkovskiy2017Making,Shchukin2024Combined} and investigated their growth kinetics\cite{Falke2020Photothermal}. 
However, implementing Raman spectroscopy entirely under UHV conditions remains uncommon. 
Current systems typically integrate optical components directly into dedicated vacuum chambers, making them complex and difficult to adapt to existing UHV infrastructure. 
To our knowledge, none of these implementations simultaneously provides precise temperature regulation, controlled gas dosing, and sample transfer within a portable platform. This gap has so far prevented systematic Raman characterization of air-sensitive GNRs across the full experimental workflow.

To address this limitation, we develop a UHV Raman platform with two modules: a portable Raman vacuum suitcase (RVS) compact enough to mount directly on the stage of a commercial Raman microscope, and a five-way transfer cross (Cross) that installs the RVS onto pre-existing UHV systems. 
The platform thereby extends established UHV infrastructure with optical spectroscopy while enabling conventional Raman systems to characterize pristine, air-sensitive materials under controlled conditions. 
The RVS minimizes surface contamination and prevents ambient exposure that would otherwise degrade the structural and electronic integrity of the GNRs. 
Samples can be grown directly under the Raman microscope, providing in situ access to reaction kinetics. 
Integrated temperature control and gas dosing further open pathways for investigating intrinsic lattice dynamics, thermal effects, and structural modifications under controlled conditions. 
We believe this versatile platform can be used to advance our understanding of low-dimensional quantum materials and their responses to environmental perturbations.

\section{Results \& discussion}

\subsection{Development of the ultra-high vacuum Raman spectroscopy platform}

\begin{figure}[!ht]
  \centering
  \includegraphics[width=\textwidth]{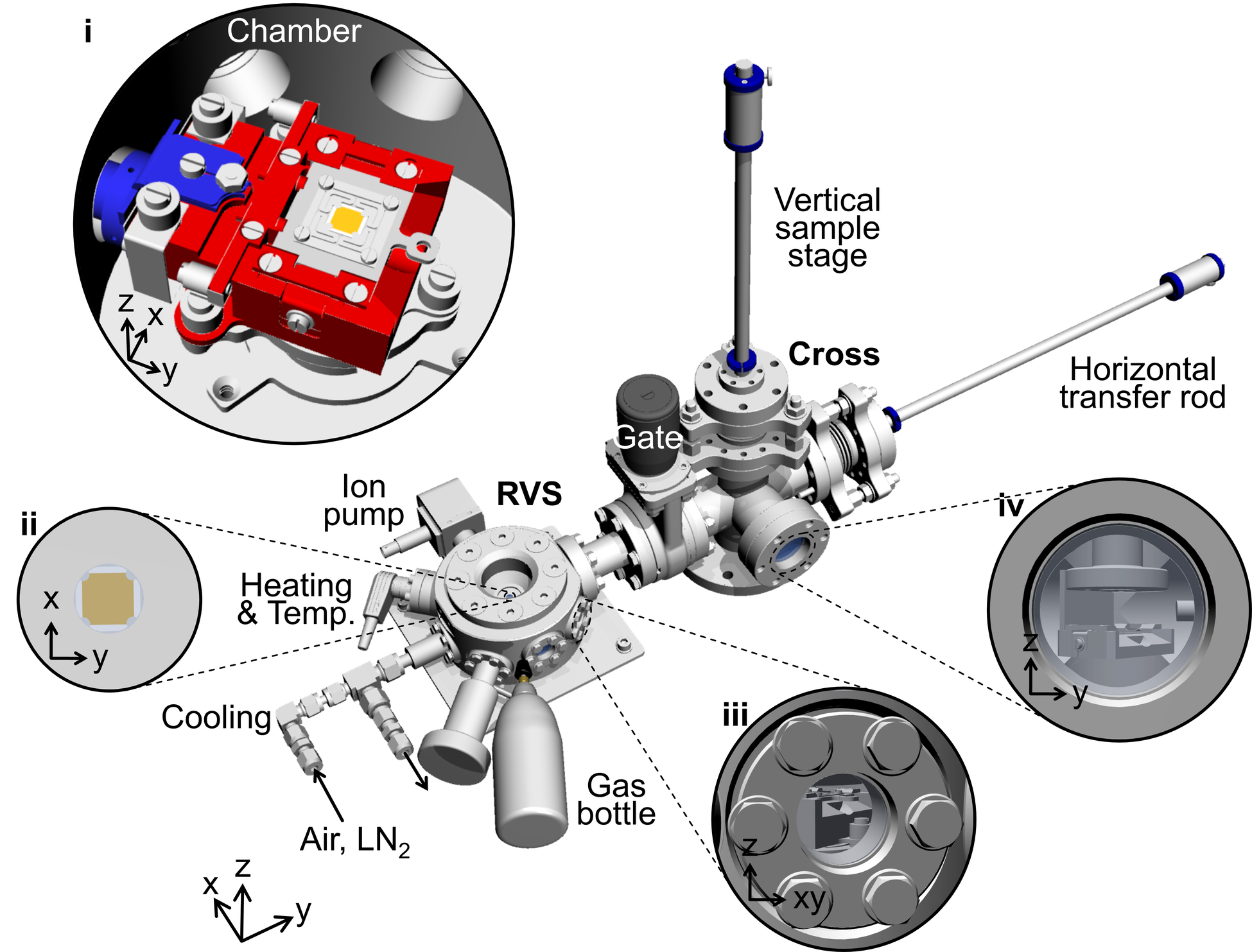}
  \caption{\textbf{Schematic illustration of the ultra-high vacuum (UHV) Raman spectroscopy platform} with the key components of the Raman vacuum suitcase (RVS) and the five-way transfer cross (Cross) depicted. Key components of the RVS: an ion pump, heating and cooling blocks with an integrated thermocouple, and a gas-dosing line equipped with a leak valve. Key components of the Cross: a vertical sample stage and a horizontal transfer rod. Insets: (i) illustration of the RVS chamber with the sample stage and cooling components highlighted in red and blue, respectively; (ii) top and (iii) side window of the RVS; (iv) front window of the Cross. Key technical specifications of the platform are summarized in Table \ref{t-spec}.}
  \label{f1}
\end{figure}

The UHV Raman platform consists of two modules separated by a gate valve: the RVS and the Cross. 
Figure \ref{f1} illustrates the main components of both units, with a photograph of the system connected to the scanning tunneling microscope (STM) preparation chamber (PC) provided in Figure \ref{fs1-1}, in which the RVS is highlighted in silver and the Cross in bronze. 
Key technical specifications of the UHV Raman platform are summarized in Table \ref{t-spec}.

\FloatBarrier

\begin{table}[!ht]
\centering
\begin{tabular}{ll}
    \hline
    Parameter & Value / Description \\
    \hline
    \multicolumn{2}{l}{\textit{Vacuum}} \\
    Base pressure & 1--3 \mul~10\super{-8} mbar \\
    Pumping & 3S TiTan Ion Pump (Gamma Vacuum, integrated) \\
    \hline
    \multicolumn{2}{l}{\textit{Temperature control}} \\
    Temperature measurement & Thermocouple (\TTC) + pyrometer (\TPy) calibration \\
    Temperature range & \TTC~= 162--748 K (-111--475 \degC) \\
    Cooling medium & Compressed air or liquid nitrogen (LN\sub{2}) \\
    Heating & Pyrolytic boron nitride (PBN) resistive heater, inert to gases \\
    \hline
    \multicolumn{2}{l}{\textit{Optical access}} \\
    Window material & Sapphire, glued with high-vacuum leak sealant \\
    Window thickness & 0.25 mm \\
    Objective & 50\mul, working distance 9.1 mm \\
    \hline
    \multicolumn{2}{l}{\textit{Gas dosing}} \\
    Gases demonstrated & \ce{O2}, \ce{N2}, \ce{CO2}, \ce{CO} \\
    Dosing range & 10\super{-8}--10\super{-5} mbar \\
    Pressure recovery & \ce{O2}: \ca 30 min to base pressure\\
        & \ce{N2}, \ce{CO2}, \ce{CO}: \ca 30 min--2 h to \ca 10\super{-7} mbar, bakeout to base pressure \\
    \hline
    \multicolumn{2}{l}{\textit{Sample transfer}} \\
    Transfer mechanism & Five-way transfer cross \\
    Connection & Fast entry lock (FEL) to STM preparation chamber \\
    Air exposure during transfer & None \\
    \hline
\end{tabular}
\caption{\textbf{Key technical specifications of the UHV Raman platform.}}
\label{t-spec}
\end{table}

To maintain UHV conditions during extended measurements and preserve sample integrity, an ion pump is integrated directly into the RVS chamber. 
Optical access to the sample is provided through a 0.25 mm-thick sapphire window positioned directly above the sample stage (highlighted in red, Figure \ref{f1}). 
The stage is equipped with a pyrolytic boron nitride (PBN) resistive heater, which is inert to the dosed gases, and a thermocouple. For temperature control, the stage is in thermal contact with a cooling block (highlighted in blue, Figure \ref{f1}). 
Its height is carefully chosen such that the sample surface lies within the 9.1 mm working distance of a 50\mul~objective, ensuring a high collection efficiency for the backscattered Raman signal (Figure \ref{fs1-1}b inset).

The RVS also enables precise temperature control and gas dosing. 
Since the stage temperature readout (\TTC) from the integrated thermocouple deviates from the surface temperature, we calibrated \TTC~against the surface temperature measured with a pyrometer (\TPy, Figure \ref{fs1-calib}). 
The calibration was performed for the two substrates most commonly used in on-surface synthesis: gold on mica (Figure \ref{fs1-calib}b) and a gold single crystal (Figure \ref{fs1-calib}c). 
For the gold single crystal, \TTC~and \TPy~show a linear relationship above \TTC~= 146 \degC, whereas the mica-supported substrate exhibits linear behavior only above \TTC~= 206 \degC. 
This difference reflects the lower thermal conductivity of mica, which produces a larger thermal gradient between the heating element and the gold surface.

Cooling is achieved via the metal plate of the cooling block, which is in thermal contact with the sample stage. 
When compressed air or liquid nitrogen (LN\sub{2}) is fed through the cooling line, heat transfer between the cooling block and the sample stage reduces the sample temperature. 
A dedicated dosing port equipped with a leak valve allows gases to be introduced into the RVS for controlled exposure experiments.

Finally, the five-way transfer cross bridges sample preparation and Raman characterization. 
Mounted directly onto the fast entry lock (FEL, Figure \ref{fs1-1}a), the Cross connects the RVS to the preparation chamber (PC) of our scanning tunneling microscope (STM). 
This configuration ensures that samples prepared in the PC can be transferred to the RVS without any exposure to ambient air. 
Together, the RVS and the Cross enable systematic, in situ Raman investigations of \pie-conjugated carbon-based nanomaterials as a function of both temperature and gas environments in UHV.

\subsection{In situ growth and Raman characterization of graphene nanoribbons}

To demonstrate in situ Raman measurements during the growth of AGNRs, we first deposited molecular precursors for 7- and 9-atom-wide AGNRs (7- and 9-AGNRs) onto clean Au(111)/mica substrates in the PC, at a base pressure of \ca 10\super{-10} mbar. 
Before performing these measurements, we confirmed that the ribbons grow correctly within the RVS itself. For this, 9-AGNR precursors were deposited onto a Au(111) surface and annealed in two-steps (\TPy~= \ca 200 and \ca 400 \degC, 10 min at each temperature). 
Figure \ref{fs2-2step-pol-gnr} shows STM images of a medium-coverage sample after each annealing step, confirming the sequential formation of polymers and then GNRs within the RVS. 

\begin{figure}[!ht]
  \centering
  \includegraphics[width=0.9\textwidth]{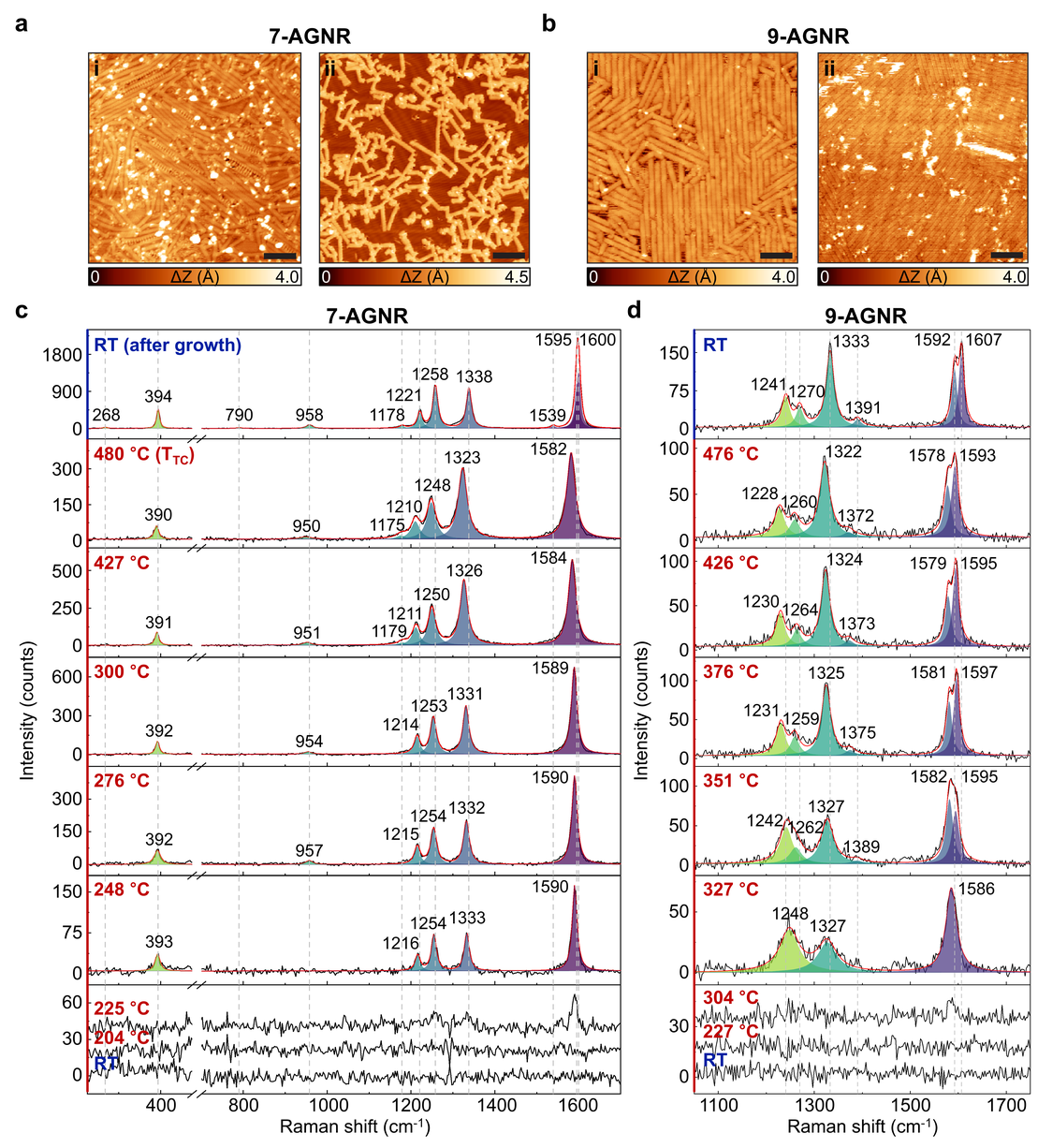}
  \caption{\textbf{In situ growth and Raman measurements on 7- and 9-AGNRs in the RVS}. Room-temperature STM images of (a) 7-AGNRs after cyclodehydrogenation in the RVS at \TTC~= (i) 427 and (ii) 480 \degC, and those of (b) 9-AGNRs after cyclodehydrogenation at \TTC~= (i) 426 and (ii) 476 \degC~(scanning parameters: -1 V, 20 pA). Scale bars: 10 nm. In-situ Raman spectra during the synthesis of (c) 7- and (d) 9-AGNRs. Initial Raman spectra were taken at room temperature (RT) following precursor deposition (the bottommost spectrum). The samples were subsequently heated in increments up to \TTC~= 480 \degC~(7-AGNR) and 476 \degC~(9-AGNR), with Raman spectra collected at each temperature step (see Figure \ref{fs2-insitu} for the complete dataset). The color-coded left axis denotes the thermal progression: the growth stages are indicated in red and the final cooldown to RT in blue. Gray dashed lines mark the final peak positions of the GNRs at RT. The spectra in the bottommost panels of (c) and (d) are stacked. Raman measurements: \excite~= 532 nm, 40 mW, 1 s integration time. At each temperature step, spectra were mapped over a 15 \um~\mul~15 \um~area (100 points) following a 10--20 min thermal stabilization, and subsequently averaged.}
  \label{f2}
\end{figure}

After verifying successful growth in the RVS, fresh precursor-covered substrates were prepared, transferred into the RVS through the Cross, and brought under the Raman microscope (Figure \ref{fs1-1}b). 
During this transfer, the samples experienced a maximum pressure of 1 \mul~10\super{-7} mbar. 
Once in the RVS, the samples were annealed stepwise up to \TTC~= 480 \degC~(7-AGNR) and 476 \degC~(9-AGNR), which correspond to calibrated surface temperatures of \ca 400 \degC. 
The resulting Raman spectra are shown in Figures \ref{f2}c and \ref{fs2-insitu}a for 7-AGNRs and Figures \ref{f2}d and \ref{fs2-insitu}b for 9-AGNRs. 
Throughout this section, we report the nominal thermocouple temperature (\TTC) rather than the calibrated surface temperature (\TPy) unless specified, as the calibration curve is linear only above \TTC~= \ca 200 \degC~(Figure \ref{fs1-calib}b).

Before analyzing the progression of the OSS, we briefly summarize the four spectral regions that contain the relevant structural information for GNRs. 
At the lowest frequencies (below \ca 150 \invcm), longitudinal compressive modes (LCMs) arise from collective vibrations of atoms along the ribbon axis\cite{Overbeck2019Universal,Hwang2025Optimized}. 
The radial breathing-like mode (RBLM) appears at a higher frequency. 
It is named after the radial breathing mode (RBM) of CNTs, whose frequency scales inversely with tube diameter\cite{Araujo2008Nature}. For GNRs, it scales inversely with the ribbon width\cite{Borin2019Surface,Liu2020In,Verzhbitskiy2016Raman,Vandescuren2008Theoretical}.

Between \ca 1100 and 1500 \invcm, GNRs exhibit Raman modes that are conventionally termed CH/D modes due to their association with \ce{C-H} bending modes and the defect-activated D band in graphene\cite{Kim2018Distinguishing,Liu2020In}. 
Recent resonance Raman studies, however, demonstrated that D modes in GNRs are intrinsic to the material, arising from phonons activated by the 1D zone-folding of the graphene dispersion\cite{Nascimento2025Optical}.

Finally, the G modes appear near 1600 \invcm, originating from in-plane \ce{C-C} stretching of the \sptwo-hybridized lattice. 
While this vibration forms a single, doubly degenerate peak in 2D graphene, the 1D geometry of GNRs breaks this degeneracy. 
It splits the G band into longitudinal optical (LO) and transverse optical (TO) components\cite{Borin2019Surface}. 
The TO mode corresponds to in-plane \ce{C-C} stretching perpendicular to the longitudinal ribbon axis, while the LO mode involves displacements parallel to this axis. 

\FloatBarrier

For the 7-AGNRs, two CH/D modes and a G mode, centered at 1254, 1333, and 1590 \invcm, first appear at 225 \degC~(calibrated to 184 \degC), marking the onset of the cyclodehydrogenation reaction. 
At 248 \degC, the RBLM (393 \invcm), CH/D modes (1216, 1254, and 1333 \invcm), and G mode (1590 \invcm) are resolved. Upon reaching 276 \degC, an additional breathing-like mode at 957 \invcm~emerges\cite{Gruneis2018Ultrahigh}, and at 328 \degC, a CH/D mode at 1182 \invcm~becomes visible (Table \ref{ts-insitu7}). 
To track these changes quantitatively, the Raman spectra at each temperature were deconvoluted into Lorentzian components guided by density functional theory (DFT) calculations (Figure \ref{fs2-normal}a). 
The evolution of peak positions and increasing intensities up to 350 \degC, followed by pronounced sharpening upon cooldown (Table \ref{ts-insitu7}), together confirm the successful growth of the 7-AGNRs. 
Note that the modes for the 7- and 9-AGNRs at 225 \degC~and 304 \degC, respectively, are visible but too weak for reliable Lorentzian fitting. 
Furthermore, anharmonic scattering and thermal expansion become prominent above 350 \degC, leading to noticeable peak broadening and redshifts of the modes\cite{Bonini2007Phonon,Klemens1966Anharmonic,Balkanski1983Anharmonic,Falke2020Photothermal}. 

The 9-AGNRs follow a similar growth trajectory. The G peak first appears at 304 \degC~(calibrated to 251 \degC), followed shortly by the CH/D modes at 327 \degC~(270 \degC). 
The onset temperature determined by Raman spectroscopy is \ca 30 \degC~lower than the 280 \degC~reported for 9-AGNRs on Au(111) by X-ray photoelectron spectroscopy (XPS)\cite{Di2018On}. 
Unlike for 7-AGNRs, the RBLM is not observed for 9-AGNRs under our experimental conditions. 
The 532 nm excitation wavelength used here lies far from the optical gap of 9-AGNRs, and does not benefit from the resonance enhancement obtained with near-infrared excitation (\excite~= 785 nm)\cite{Borin2019Surface}. 
Following the procedure used for the 7-AGNRs, the 9-AGNR spectra were deconvoluted using calculated Raman modes (Figure \ref{fs2-normal}b). 
The evolution of the constituent peaks (Figures \ref{f2}d, \ref{fs2-insitu}b, Table \ref{ts-insitu9}) tracks the formation of the ribbons.

After cooling to room temperature, the 7-AGNRs exhibit a shear-like mode (SLM) at 268 \invcm~and well-resolved G modes at 1595 (LO) and 1600 \invcm~(TO). 
For both ribbons, the Raman spectra of the samples grown in the RVS collected after cooldown agree with those of equivalent samples prepared in the standard STM preparation chamber (Figure \ref{fs2-Raman_InSitu-vs-vt}). 
This demonstrates that the RVS-based growth yields comparable structural quality.

An interesting feature in the Raman spectra of the 9-AGNRs is the pronounced intensity of the TO mode relative to the LO mode in the G mode region. 
This behavior is consistent with the excitation-dependent Raman response of 9-AGNRs, in which the TO mode is preferentially enhanced at 532 nm\cite{Borin2019Surface, Nascimento2025Optical}. 
For the 7-AGNRs, the LO (1595 \invcm) and TO (1600 \invcm) modes become discernible only after cooldown because the separation (\ca 5 \invcm) is smaller than that of the 9-AGNRs (\ca 15 \invcm; 1592 and 1607 \invcm). 
These peak positions align well with previously reported values for 7- and 9-AGNRs on \ce{SiO2}/\ce{Si} substrates\cite{Borin2019Surface}.

STM images acquired after growth in the RVS further validate the structural quality of the synthesized ribbons. 
For the 7-AGNRs, end-to-end fusion occurred when annealing reached 480 \degC~(Figure \ref{f2}a). 
We attribute this to thermally activated coupling of carbon radicals at the 7-AGNR termini\cite{Talirz2013Termini}, possibly favored by the elevated pressure in the RVS during growth (\ca 10\super{-6} mbar) compared with standard OSS conditions (\ca 1--10 \mul~10\super{-10} mbar)\cite{Fairbrother2017High}. 
In contrast, the 9-AGNRs show no noticeable fusion, regardless of whether growth is terminated at 426 or 476 \degC~(Figure \ref{f2}b).

To further quantify the effect of chamber pressure on GNR quality, the 9-AGNR precursors were annealed in the RVS under three distinct pressure environments (Figure \ref{fs2-pressure}). 
By introducing air cooling and opening the gate valve to the FEL turbo pump, the chamber pressure was modulated to create three growth conditions: (a) gate closed, no air cooling, (b) gate closed, with air cooling, and (c) gate open, with air cooling. 
Air cooling reduces outgassing from the heated RVS, while the open gate introduces additional pumping capacity from the FEL turbo pump. 
We find that the final ribbon length is inversely correlated with the pressure during the growth (Figures \ref{fs2-pressure}e, \ref{fs2-hist}). 
The average length increases from 11.5 \plm~4.6 nm for the highest-pressure condition (a, 9.2 \mul~10\super{-7}--6.8 \mul~10\super{-6} mbar) to 34.6 \plm~12.6 nm (b, 1.7 \mul~10\super{-7}--8.4 \mul~10\super{-7} mbar) and 47.5 \plm~12.7 nm for the lowest-pressure condition (c, 6.1 \mul~10\super{-8}--1.2 \mul~10\super{-7} mbar). 
As a reference, 9-AGNRs grown in the PC at 2 \mul~10\super{-10} to 1.5 \mul~10\super{-9} mbar have an average length of 74.1 \plm~25.0 nm. 
These results confirm that minimizing pressure during OSS yields significantly longer GNRs, consistent with the study by Fairbrother et al.\cite{Fairbrother2017High}.

\subsection{Temperature-dependent Raman spectroscopy}
\begin{figure}[!ht]
  \centering
  \includegraphics[width=0.7\textwidth]{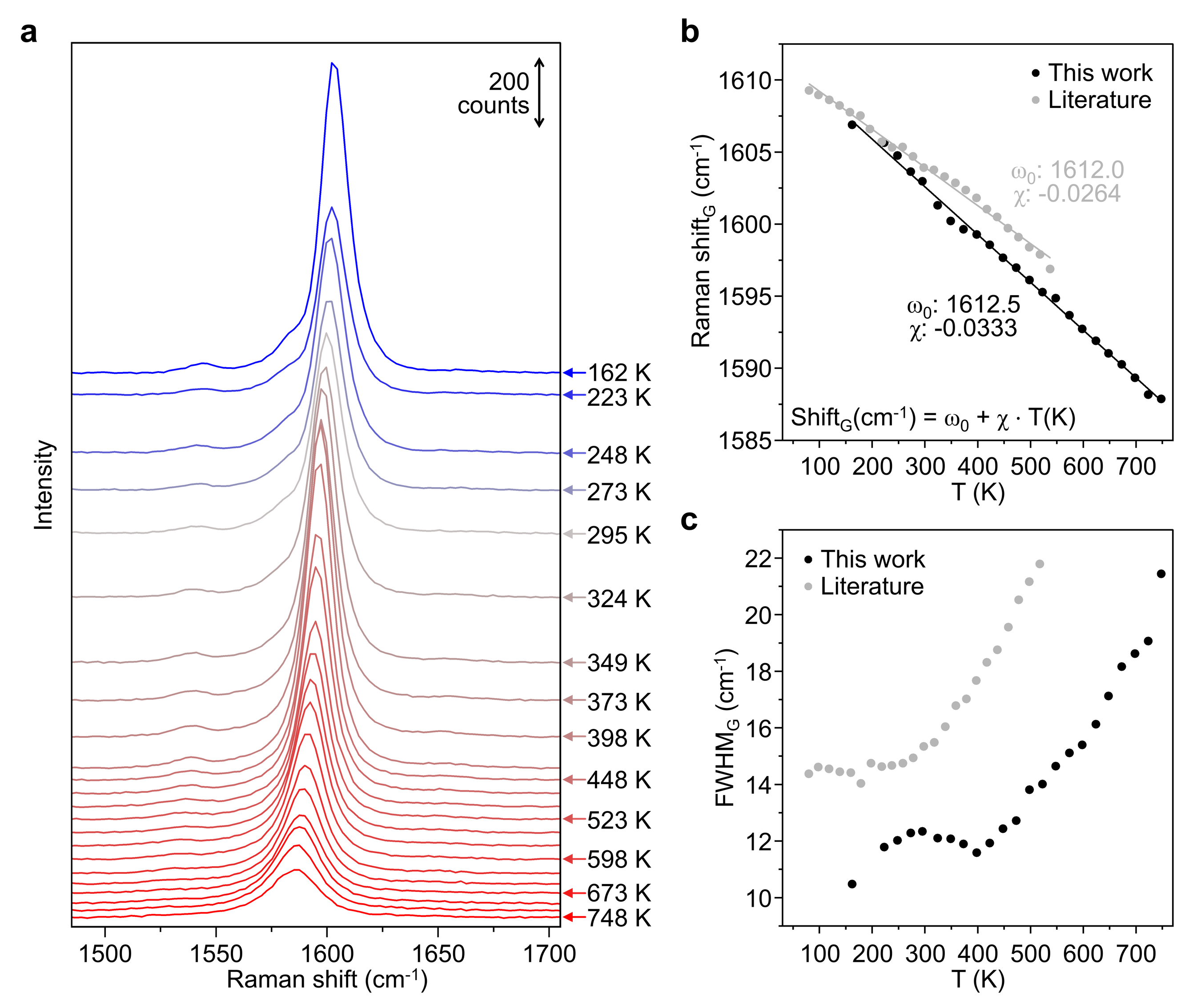}
  \caption{\textbf{Temperature dependence of the G peak.} (a) Spectral evolution of the G mode with temperature. Evolution of the (b) peak position and (c) full width at half maximum (FWHM) of the G mode as a function of temperature. Black data points are the values extracted in this study, and gray data points are those reported by Guo et al.\cite{Guo2022Phonon}. Raman measurement conditions: \excite~= 532 nm, 40 mW, 1 s. Spectra were mapped over a 15 \um~\mul~15 \um~area (100 points) following thermal stabilization and subsequently averaged. Full spectra and detailed peak information are provided in Figure \ref{fs3-1} and Table \ref{ts-tdep7}, respectively.}
  \label{f3}
\end{figure}

Temperature-dependent Raman spectroscopy is a key method for probing fundamental physical properties of low-dimensional materials, and RVS gives direct access to it. 
Accessible properties include phonon anharmonicity\cite{Calizo2007Temperature,Guo2022Phonon}, electron--phonon coupling\cite{Pisana2007Breakdown,Piscanec2004Kohn}, thermal conductivity\cite{Balandin2008Superior,Ghosh2010Dimensional,Cai2010Thermal}, and thermal expansion\cite{Bao2009Controlled,Yoon2011Negative}. 
To demonstrate this capability, we prepared a pristine 7-AGNR sample in the STM PC and collected Raman spectra over a temperature range from \TTC~= 162--748 K (-111--475 \degC). 
Figure \ref{fs3-1} presents the evolution of the full spectra, and Figure \ref{f3}a that of the G-mode region. Cooling was achieved using LN\sub{2} with counter-heating. 
All temperatures reported in this section correspond to the thermocouple readings.

\FloatBarrier

As a benchmark, we compare our temperature-dependent Raman data for 7-AGNRs with those reported by Guo et al. (gray data points, 7-AGNRs on Au(111), 80--520 K)\cite{Guo2022Phonon}. 
In both cases, the Raman modes shift systematically with temperature, reflecting the underlying lattice anharmonicity and phonon--phonon interactions\cite{Poulos2024Temperature, Bonini2007Phonon}. 
To quantify this behavior, we linearly fitted the peak positions ($\omega$) as a function of temperature for the CH/D and G modes. 
From these fits, we extracted the first-order temperature coefficient ($\chi$, the slope) and the Raman shift extrapolated to 0 K ($\omega_0$). 

The extracted $\chi$ and $\omega_0$ values for the three CH/D modes and the TO component of the G mode are summarized in Table \ref{t1} together with the literature values\cite{Guo2022Phonon} (Figures \ref{f3}b, \ref{fs3-trend}, Table \ref{ts-tdep7}). 
The $\omega_0$ values agree within 4 \invcm, and the temperature coefficients have the same sign and magnitude (deviations of 10--35\%), which is reasonable given the different measurement conditions. 
This demonstrates reliable temperature control within the RVS and confirms the quality of the acquired spectra. 

\begin{table}[!ht]
\centering
\begin{tabular}{c c c c c c}
\hline
         &  Value &   \multicolumn{3}{c}{CH/D}   & G\\ \hline
\multirow{2}{*}{This work}   & $\omega$\sub{0} & 1228.6 & 1266.6 & 1348.9 & 1612.5\\ 
            & $\chi$ & -0.0254  & -0.0243  & -0.0301  & -0.0333\\
            \hline
\multirow{2}{*}{Guo et al.\cite{Guo2022Phonon}}  & $\omega$\sub{0} & 1229.9  & 1266.5  & 1352.3 & 1612.0\\ 
            & $\chi$ & -0.023  & -0.018  & -0.0307  & -0.0264\\
            \hline
\end{tabular}
\caption{\textbf{Raman shifts at 0 K ($\omega$\sub{0}) and first-order temperature coefficients ($\chi$) of the CH/D and G modes} from this work compared to those reported by Guo et al.\cite{Guo2022Phonon} for 7-AGNRs on Au(111). The units are \invcm~for $\omega$\sub{0} and \invcm/K for $\chi$.} 
\label{t1}
\end{table}
\FloatBarrier

Figures \ref{f3}b,c and \ref{fs3-trend} show the corresponding temperature dependence of the RBLM, CH/D, and G mode positions and linewidths. 
The RBLM shows a nonlinear redshift accompanied by an increase in its full width at half maximum (FWHM) with increasing temperature (Figure \ref{fs3-trend}a). 
In contrast, the CH/D and G modes show a predominantly linear redshift with temperature (Figures \ref{fs3-trend}b and \ref{f3}), and an overall increase in their FWHM. 
These trends are in good agreement with the reported temperature dependence of the Raman modes of 7-AGNRs\cite{Guo2022Phonon}, and allow us to probe lattice phonon anharmonicity and phonon--phonon decay\cite{Poulos2024Temperature, Bonini2007Phonon,Klemens1966Anharmonic,Balkanski1983Anharmonic}.

The Raman spectra of the 7-AGNRs measured at room temperature after different thermal treatments (Figure \ref{fs3-after-diff-temp-trend}) do not reveal critical structural degradation. 
Our data, however, differ from the literature in two notable aspects. 
First, the measured peak widths are narrower on average. 
Unlike the samples of Guo et al. that were exposed to ambient conditions before the measurement\cite{Guo2022Phonon}, the 7-AGNRs in this study were strictly maintained in vacuum between growth and measurement, ensuring sample cleanliness. 
Consequently, scattering from surface adsorbates is minimized and intrinsic phonon lifetimes are preserved\cite{Bonini2007Phonon,Ferrari2013Raman}. 
Second, the FWHM of the CH/D and G modes exhibits a non-monotonic evolution with two noticeable deviations near 300 K and 400 K from the overall increasing trend. 
This behavior likely reflects a complex interplay among competing scattering mechanisms, including temperature-dependent electron--phonon and phonon--phonon coupling, substrate interactions, and multi-phonon processes\cite{Han2022Raman,Kolesov2017Low,Menendez1984Temperature,Guo2022Substrate}. 
Disentangling these contributions will require further experiments, such as measurements on different substrates and multi-wavelength Raman spectroscopy. 

\subsection{Controlled gas dosing}

Recent scanning probe microscopy studies under UHV have provided direct evidence of chemical instability at zigzag edge segments. 
Bond-resolved STM has imaged the oxidation of zigzag termini in 5-AGNRs\cite{Lawrence2020Probing} and of the zigzag segments of (3,1)-chiral GNRs\cite{Lawrence2022Circumventing,Berdonces2021Chemical}, showing that these reactive sites readily form ketone or other oxygenated groups upon oxygen exposure. 
While such measurements offer atomic-scale insights, they are largely restricted to metallic substrates and do not transfer easily to device-relevant substrates. 
A method capable of detecting chemical changes across different substrates and after device integration is therefore highly desirable. 
The RVS allows us to study material reactivity and stability under well-defined gas environments using Raman spectroscopy.

\begin{figure}[!ht]
  \centering
  \includegraphics[width=\textwidth]{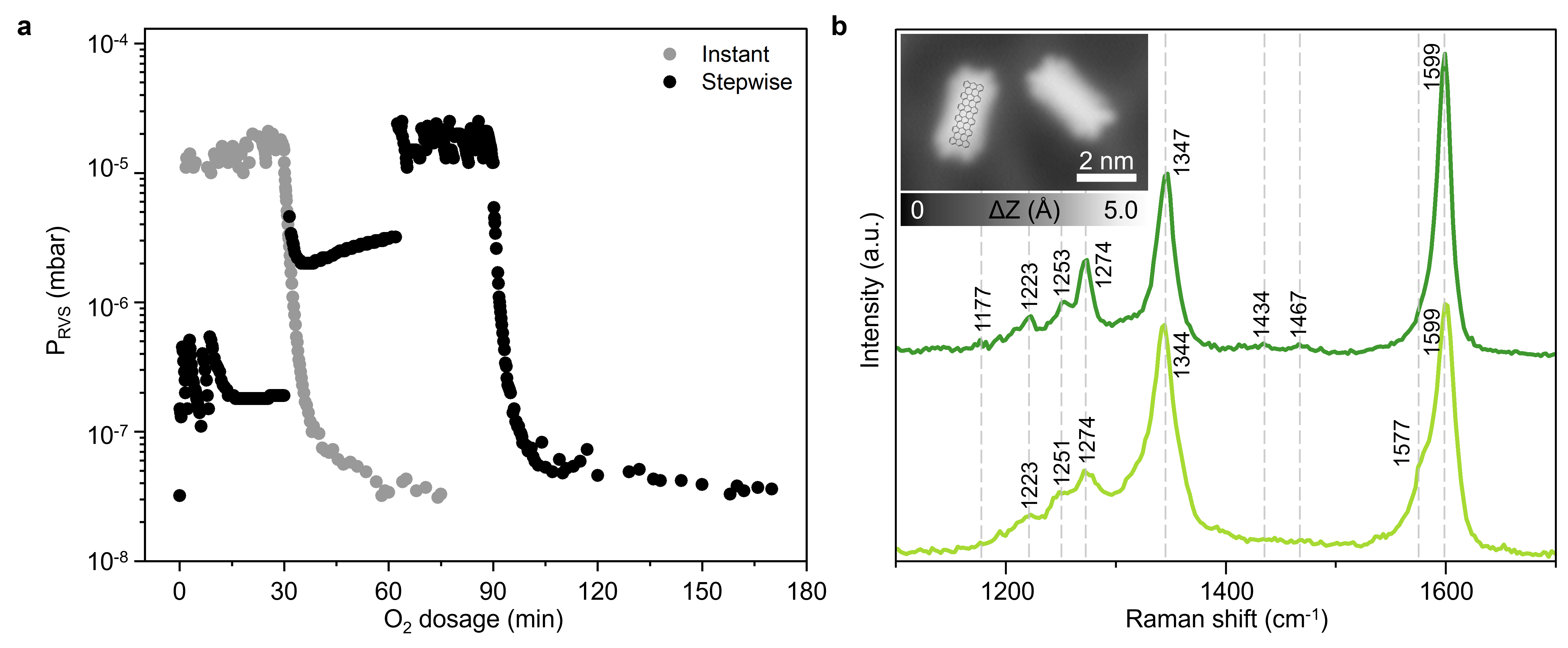}
  \caption{\textbf{Controlled gas dosing in the vacuum suitcase.} (a) Pressure curves during stepwise (black) and single-step (gray) \ce{O2} dosing up to the 10\super{-5} mbar range. (b) Raman spectra of short 7-AGNRs collected before (green) and after (light green) oxygen exposure. All spectra were acquired over a 30 \um~\mul~30 \um~area (400 points, \excite~= 532 nm, 20 mW, 5 s integration time). Inset: STM image of short 7-AGNRs (scanning parameters: -100 mV, 30 pA, 4.4 K).}  
  \label{f4}
\end{figure}
\FloatBarrier

To assess the dosing and pumping behavior of gases with different reactivities, we introduced \ce{O2}, \ce{N2}, \ce{CO2}, and \ce{CO} into the RVS chamber. 
Starting from a base pressure of 1--3 \mul~10\super{-8} mbar, gases can be dosed against the running ion pump up to the 10\super{-5} mbar range. 
For \ce{O2} (Figure \ref{f4}a), the pressure in the RVS (P\sub{RVS}) recovers to base pressure within \ca 30 min, regardless of whether the pressure was raised in a single step (gray trace) or stepwise (black trace). 
In contrast, the pressure recovery is far slower when \ce{N2}, \ce{CO2}, and \ce{CO}, which are pumped less efficiently than \ce{O2}, are dosed (Figure \ref{fs4-gas}). 
P\sub{RVS} reaches the 10\super{-7} mbar range after \ca 30 min--2 h, and full recovery to base pressure requires a bakeout. 
This behavior suggests that extended dosing of these gases at \ca 1--10 \mul~10\super{-6} mbar compromises the ion pump performance, likely reflecting cathode saturation and re-emission of weakly bound species\cite{Audi1987Ion}.

We then used the dosing capability of the RVS to probe the overall chemical reactivity of the three ribbon systems sharing the 7-AGNR backbone by Raman spectroscopy. 
First, short 7-AGNRs were selected to probe oxidation at zigzag edges. 
Their extremely short length (1--3 nm) results in a high zigzag-to-armchair edge ratio, and they are particularly reactive at the zigzag termini despite being armchair ribbons\cite{Borin2023On}. 
Figure~\ref{f4}b shows Raman spectra of short 7-AGNRs before (green) and after (light green) \ce{O2} exposure. 
Under 532 nm excitation, the \ce{C-H} bending mode at 1177 \invcm~that involves strong displacement of the H atoms at the center carbons of the zigzag termini\cite{Borin2023On}, and the vibrations involving edge C and H atoms at 1434 and 1467 \invcm~are no longer observed after dosing (Figure \ref{fs4-s7_pr_norm}). 

The spectral changes are even more pronounced with 488 nm excitation (Figure \ref{fs4-monofull}). 
Two \ce{C-H} bending modes are resolved before exposure, at 1138 and 1169 \invcm, which correspond to strong hydrogen displacements at the side and center carbons of the zigzag termini, respectively (Figure \ref{fs4-s7_pr_norm})\cite{Borin2023On}. 
Both modes disappear after \ce{O2} exposure (Figures \ref{fs4-s7_pr_norm}, \ref{fs4-monobr_dft}, Table \ref{ts-o2-s7}), indicating chemical modification at the zigzag termini. 
Previous UHV-STM studies have demonstrated the chemical sensitivity of (3,1)-chiral GNR zigzag motifs\cite{Lawrence2022Circumventing,Berdonces2021Chemical}. 
Based on the reported chemical changes, the center carbon atoms of short 7-AGNR zigzag termini may undergo hydrogenation, ring contraction, introduction of \ce{=O} or \ce{-OH} groups, and formation of \ce{C-O-C}. The oxidation scenarios with \ce{-OH} and \ce{=O} formation are considered in our calculations (Figure \ref{fs4-monobr_dft}). 
See Supporting Information (Figures \ref{fs4-s7_pr_norm}, \ref{fs4-monobr_dft}, \ref{fs4-s7_o2_norm}) for a detailed discussion of the calculated normal modes and spectral changes.

Analogous dosing experiments were carried out for conventional 7-AGNRs and staggered 7-AGNR-S(1,3) (Figure \ref{fs4-topoO2}). 
The latter are referred to as topological GNRs. 
For 7-AGNRs, which are known to be stable under ambient conditions, spectral changes after \ce{O2} dosing are negligible. 
All characteristic Raman peaks remain present, and their position differences are below the experimental resolution (\ca 3 \invcm). 
In contrast, topological GNRs, whose topologically protected in-gap states substantially reduce the electronic bandgap relative to pristine 7-AGNRs\cite{Groning2018Engineering}, show pronounced spectral changes upon \ce{O2} exposure. 
Although the majority of the modes remain observable, their shifts, broadening, and the disappearance of several modes indicate that the ribbons no longer retain their pristine form (Figures \ref{fs4-topoO2}--\ref{fs4-topo_o2_norm}, Table \ref{ts-o2-topo}). 
The structural imperfections of the present sample (Figure \ref{fs4-topoO2}c) likely contribute to its reduced stability, despite previous reports of air stability on Au(111)\cite{Groning2018Engineering} and degradation only upon substrate transfer\cite{Kinikar2025Atomic}. 
Collectively, these results demonstrate that the UHV Raman workflow using the RVS can sensitively detect chemical modifications of GNRs, including edge-specific modifications, under controlled gas environments.

\section{Conclusion}
In conclusion, we have introduced a compact, portable Raman vacuum suitcase as a versatile platform for the characterization of atomically precise GNRs under UHV conditions. 
The RVS can be installed on conventional UHV chambers and mounted directly on the stage of a commercial Raman microscope, extending established infrastructure with in situ optical spectroscopy. 
Using this setup, we tracked the on-surface synthesis of 7- and 9-AGNRs, monitored the temperature-dependent evolution of 7-AGNR Raman modes from 162 to 748 K, and resolved chemical changes in topological and short 7-AGNRs under controlled \ce{O2} exposure. 
The in situ Raman measurements capture the emergence of vibrational fingerprints during on-surface synthesis, providing direct optical access to growth kinetics. 
The temperature-dependent measurements yielded first-order temperature coefficients consistent with literature values, enabling quantitative studies of phonon anharmonicity and lattice dynamics in pristine ribbons without ambient contamination. 
For short 7-AGNRs, controlled \ce{O2} exposure resulted in the selective disappearance of \ce{C-H} bending modes at the zigzag termini, consistent with oxidation at reactive zigzag sites. 
The dosing of topological GNRs further revealed systematic shifts and broadening of intrinsic Raman modes, pointing to structural modification of the ribbons. 
These findings demonstrate the sensitivity of Raman spectroscopy in UHV to chemical modifications. 
Owing to its portability and compatibility with conventional UHV and Raman infrastructure, the RVS provides a practical bridge between on-surface synthesis and device integration, where tracking structural integrity during substrate transfer remains a central challenge. 
We anticipate that this methodology will extend to other reactive low-dimensional materials and enable deeper insight into their fundamental properties and responses to thermal and chemical perturbations.

\section{Methods}
\textbf{\textit{Precursor synthesis}} 
The 10,10′-dibromo-9,9′-bianthryl (\textbf{DBBA})\cite{Cai2010Atomically}, 3′,6′-diiodo-1,1′:2′,1″-terphenyl (\textbf{DITP})\cite{Di2018On}, 6,11-bis(10-bromoanthracen-9-yl)-1,4-dimethyltetracene (\textbf{BADMT})\cite{Groning2018Engineering}, and 10-bromo-9,9′:10′,9″-teranthracene (\textbf{BTA})\cite{Borin2023On}, respectively for 7-AGNRs, 9-AGNRs, topological GNRs (7-AGNR-S(1,3)), and short 7-AGNRs, were synthesized according to previously reported procedures. 
The identity and purity of the final products were confirmed by nuclear magnetic resonance (NMR) analysis.

\textbf{\textit{GNR preparation}} 
Clean Au(111) surfaces were prepared from Au(111) single crystals (MaTeck GmbH) or 200-nm-thick Au(111) films on mica (4 \mul~4 mm\super{2}, PHASIS S\`arl) by iterative Ar\super{+} sputtering and annealing cycles. 
To grow 7-AGNRs, the \textbf{DBBA} molecules were loaded into the quartz crucible of a home-built evaporator and sublimed at 195 \degC~onto a clean Au(111) substrate at room temperature. 
After deposition, the sample was annealed at 200 \degC~for 10 min to induce dehalogenative polymerization, followed by an additional 10-minute annealing at 400 \degC~for cyclodehydrogenation. 
To grow 9-AGNRs, the \textbf{DITP} molecules were sublimed at 65 \degC~and annealed at 200 \degC~and then at 400 \degC, maintaining each temperature for 10 min. 
Topological GNRs were grown by subliming \textbf{BADMT} molecules at 370 \degC, followed by consecutive 10-minute annealing steps at 200, 255, 280, and 350 \degC. 
Finally, \textbf{BTA} molecules were sublimed at 370 \degC~and annealed at 265 \degC~for 10 min to grow short 7-AGNRs. 
For the in situ growth of 7- and 9-AGNRs, \textbf{DBBA} (for 7-AGNRs) and \textbf{DITP} (for 9-AGNRs) precursor molecules were first deposited onto clean Au(111) surfaces. 
The samples were then transferred to the RVS via the FEL. 
Annealing was performed by increasing \TTC~in increments of \TTC~= \ca 25 \degC, with 10--20 min of temperature stabilization at each step.

\textbf{\textit{Gas dosing}} 
The base pressure of the RVS was maintained at \ca 1--3 \mul~10\super{-8} mbar. Target gases, including \ce{O2}, \ce{N2}, \ce{CO2}, and \ce{CO}, were dosed up to \ca 3 \mul~10\super{-5} mbar. 
In the event of pump saturation following the introduction of \ce{N2}, \ce{CO2}, and \ce{CO}, the initial base pressure was restored by an overnight bakeout at \ca 100 \degC.

\textbf{\textit{STM characterization}} 
STM measurements were performed with commercial variable-temperature and low-temperature (4.4 K) STM from Scienta Omicron at base pressures below 5 \mul~10\super{-11} mbar. 
STM images were acquired in constant-current mode. 
All data were processed with WaveMetrics Igor Pro.

\textbf{\textit{Raman spectroscopy}} 
Raman spectra were measured in backscattering geometry using a WITec Alpha 300 R confocal Raman microscope, with a 600 g/mm grating for \excite~= 488 and 532 nm. 
All spectra were collected using a 50\mul~LD objective (Zeiss, NA = 0.55, 9.1 mm working distance). 
For all post-dosing measurements, the RVS was allowed to recover to base pressure before data acquisition. 
The Raman spectra were fitted using Lorentzian profiles. 
The FWHM of neighboring modes was constrained to a shared value. All data were processed with OriginPro. 

\textbf{\textit{DFT calculations of Raman spectra and normal modes}} 
Finite DFT calculations were employed to simulate the Raman spectra of short 7-AGNRs in their pristine form and functionalized with \ce{-OH} and \ce{=O} groups, and periodic calculations were used for 7- and 9-AGNRs, along with the topological GNRs (pristine and \ce{-OH}- and \ce{=O}- functionalized).

\textit{Finite molecular calculations} were performed with the ORCA 5.0.4 DFT code\cite{Neese2011ORCA} using the Perdew--Burke--Ernzerhof (PBE) exchange-correlation functional\cite{perdew_Generalized_1996} with the def2-SVP basis set and the RI-J approximation for the Coulomb integrals. 
Vibrational frequencies were computed numerically. 
Atomic positions were relaxed prior to the Raman calculations using the same settings and the default convergence threshold.

\textit{Periodic calculations} were carried out using AiiDAlab-QE, a Q\sub{UANTUM} ESPRESSO application\cite{Wang2026Making} in combination with the vibroscopy external plugin\cite{Bastonero2024Automated}. 
This plugin integrates automated AiiDA-based workflows for phonon and Raman calculations by employing a finite-displacement and finite-field approach. 
The DFT calculations were performed using Q\sub{UANTUM} ESPRESSO 7.4\cite{Giannozzi2020}. 
The nanoribbons were aligned along the $x$ axis. To simulate isolated 1D ribbons, vacuum spacings of 14 \AA~along $y$ and 25 \AA~along $z$ were added to the ribbon dimensions to avoid spurious interactions between periodic images. 
Before computing the vibrational spectra, the atomic positions and the cell parameter along the $\mathbf{a}$ vector were optimized using the PBE functional until the forces on all atoms fell below 5 \mul~10\super{-5} Ry\,Bohr\super{-1}, with an SCF convergence threshold of 1 \mul~10\super{-15} Ry. 
Norm-conserving pseudopotentials from the PseudoDojo library (stringent, v0.4) were employed\cite{vanSetten2018}. 
A plane-wave cutoff of 92.0 Ry was used for the wavefunctions, while a cutoff of 368.0 Ry was applied for the charge density. The SCF calculations used fixed occupations. 
Because the ribbons are periodic only along $x$, the Brillouin zone was sampled with a $k$-point grid resolution of 0.08 \AA\super{-1} along this direction and a single $k$-point (1 \mul~1) along the non-periodic $y$ and $z$ directions.

\section*{Author contribution}
G.B.B., A.K., P.R., and R.F. conceived the project. J.H.H., A.K., and L.R. developed the RVS. J.H.H. performed the experiments and analyzed the data. J.H.H., M.L.P., A.O.-G., and C.A.P. conducted the DFT Raman simulations. K.M., T.G.E., and X.F. synthesized the molecular precursors. J.H.H. wrote the manuscript with input from all co-authors. G.B.B. supervised the project. G.B.B., M.L.P., and R.F. acquired funding and revised the manuscript.

\section*{Statement of conflict of interest}
The authors declare no competing interests.

\section*{Supporting Information}
The Supporting Information is available online: photographs and technical drawing of the RVS; thermocouple calibration; STM images of 9-AGNRs after two-step growth in the RVS; in situ Raman spectra and Lorentzian fits during growth; DFT normal modes of 7- and 9-AGNRs; comparison of GNRs grown in the RVS and in the PC; pressure dependence of GNR length and length histograms; temperature-dependent Raman spectra, peak positions, and linewidths of 7-AGNRs; Raman spectra after thermal cycling; pressure recovery after dosing \ce{N2}, \ce{CO2}, and \ce{CO}; calculated Raman spectra and normal modes of pristine and oxidized short 7-AGNRs and topological GNRs; Raman spectra of 7-AGNRs and topological GNRs before and after \ce{O2} exposure; tables of fitted peak parameters.

\section*{Acknowledgements}
J.H.H., A.K., L.R., A.O.-G., C.A.P., R.F., and G.B.B. acknowledge funding from the Werner Siemens Foundation (CarboQuant). 
M.L.P. acknowledges funding from the Swiss National Science Foundation under the Eccellenza Professorial Fellowship no. PCEFP2\textunderscore203663 and project number 10.004.856, as well as support from the Swiss State Secretariat for Education, Research and Innovation (SERI) under contract number MB22.00076 (ERC Starting Grant E-CONVERT). 
G.B.B. and R.F. acknowledge funding from the Swiss National Science Foundation under grant no. 212875 (SYNC) and the European Union Horizon 2020 research and innovation program under grant agreement no. 881603 (Graphene Flagship Core 3). 
G.B.B., R.F., and M.L.P. acknowledge funding from the European Union’s Horizon Europe research and innovation program under grant agreement no. 101099098 (ATYPIQUAL).  G.B.B. acknowledges funding from the Swiss National Science Foundation grant no. 200021E-219172/1 (GRAAL). 
A.O.-G. and C.A.P. acknowledge support from the NCCR MARVEL, a National Centre of Competence in Research, funded by the Swiss National Science Foundation (Grant number 205602). 
DFT calculations were performed on the Swiss National Supercomputing Centre (CSCS) supercomputer (project ID: lp83) and on the ETH Z\"urich (Euler) supercomputer. 
The authors thank Dr. Oliver Braun and Dr. Jan Overbeck for technical discussions.

\begin{singlespace}
\raggedright
\bibliographystyle{unsrtnat}
\bibliography{ref}
\end{singlespace}

\clearpage

\section*{Supporting Information}
\setcounter{figure}{0}
\setcounter{table}{0}
\renewcommand{\thefigure}{S\arabic{figure}}
\renewcommand{\thetable}{S\arabic{table}}

\begin{figure}[!ht]
    \centering
    \includegraphics[width=\textwidth]{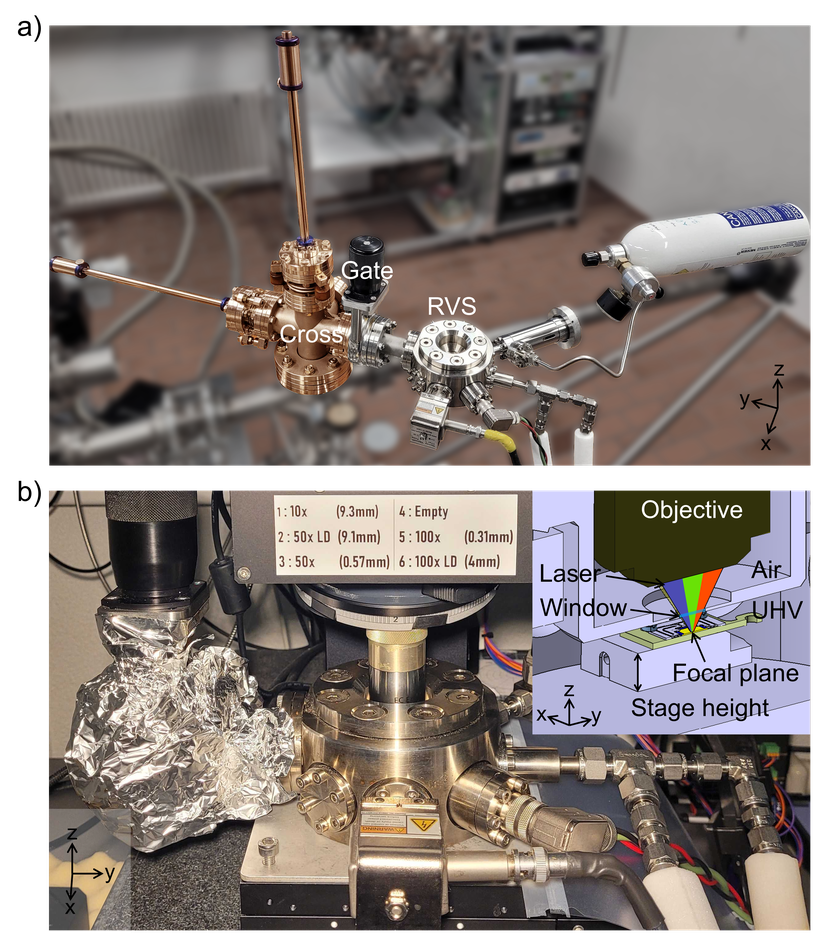}
    \caption{\textbf{Photographs and technical illustration of the RVS.} (a) Photograph of the RVS (silver) mounted to the FEL via the Cross (bronze). (b) A photograph of the RVS positioned under the Raman microscope. The inset shows a cross-sectional illustration of the RVS, demonstrating how the microscope objective accesses the sample surface through the top window.}
    \label{fs1-1}
\end{figure}
\FloatBarrier

\begin{figure}[!ht]
    \centering
    \includegraphics[width=0.8\textwidth]{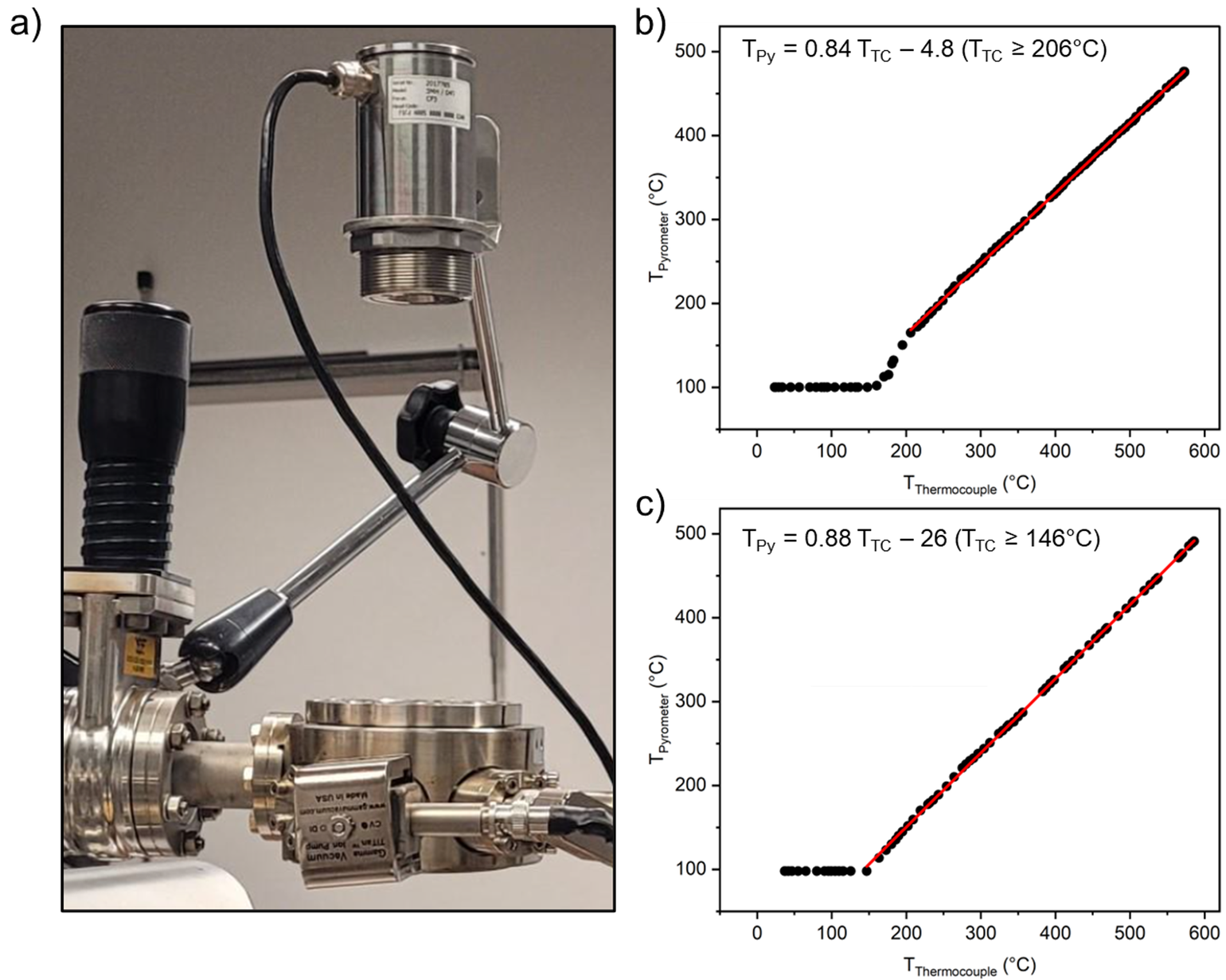}
    \caption{\textbf{Calibration of the RVS thermocouple temperature readout}. (a) Calibration setup. An optical pyrometer was used to measure the surface temperature of the sample, providing a reference for the thermocouple readout. Calibration curves for (b) Au/mica and (c) Au(111) single crystal.}
    \label{fs1-calib}
\end{figure}
\FloatBarrier

\begin{figure}[!ht]
    \centering
    \includegraphics[width=0.7\textwidth]{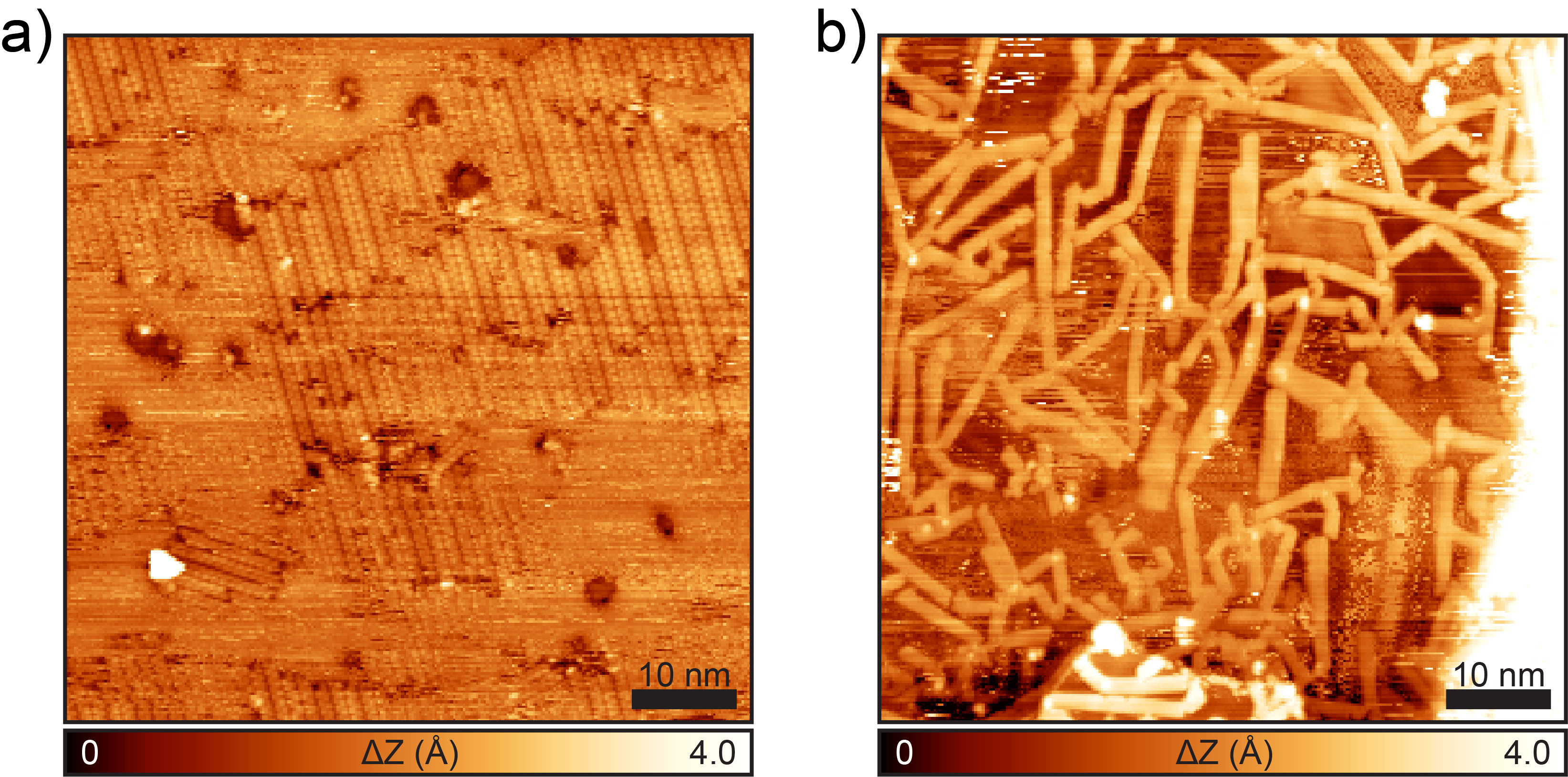}
    \caption{\textbf{STM images of 9-AGNRs after two-step growth in the RVS.} STM images of (a) polymers after annealing at \TPy~= \ca 200 \degC~for 10 min and (b) fully cyclodehydrogenated 9-AGNRs after subsequent annealing at \TPy~= \ca 400 \degC~for 10 min. All images were acquired at room temperature. Scanning parameters: -1 V, 30 pA.}
    \label{fs2-2step-pol-gnr}
\end{figure}
\FloatBarrier

\begin{sidewaysfigure}[!ht]
    \centering
    \includegraphics[width=\textwidth]{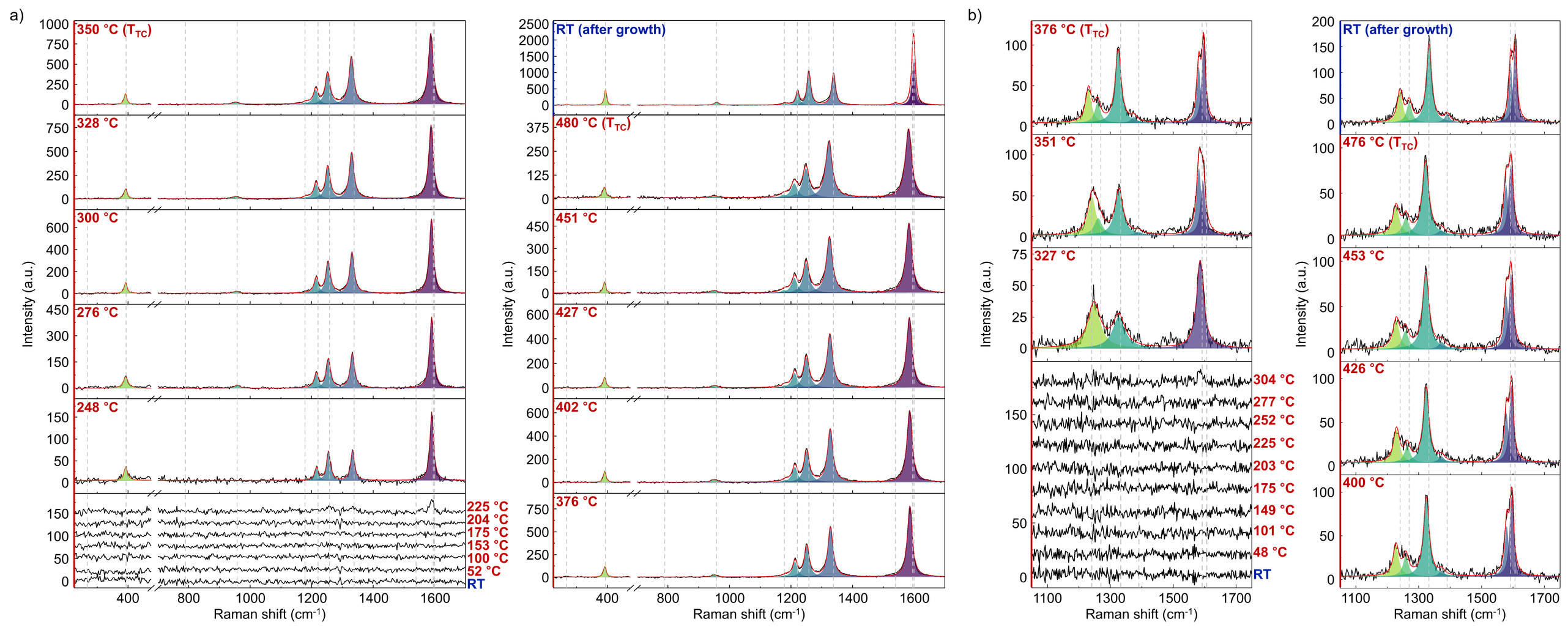}
    \caption{\textbf{Raman spectra acquired during and after the OSS in the RVS.} Raman spectra of (a) 7- and (b) 9-AGNRs. Experimental spectra are shown as black lines, fitted curves as red lines, and Lorentzian components as color-filled peaks. Gray dashed lines indicate peak positions at room temperature. Spectra in the bottommost panels are stacked.}
    \label{fs2-insitu}
\end{sidewaysfigure}
\FloatBarrier

\begin{figure}[!ht]
    \centering
    \includegraphics[width=\textwidth]{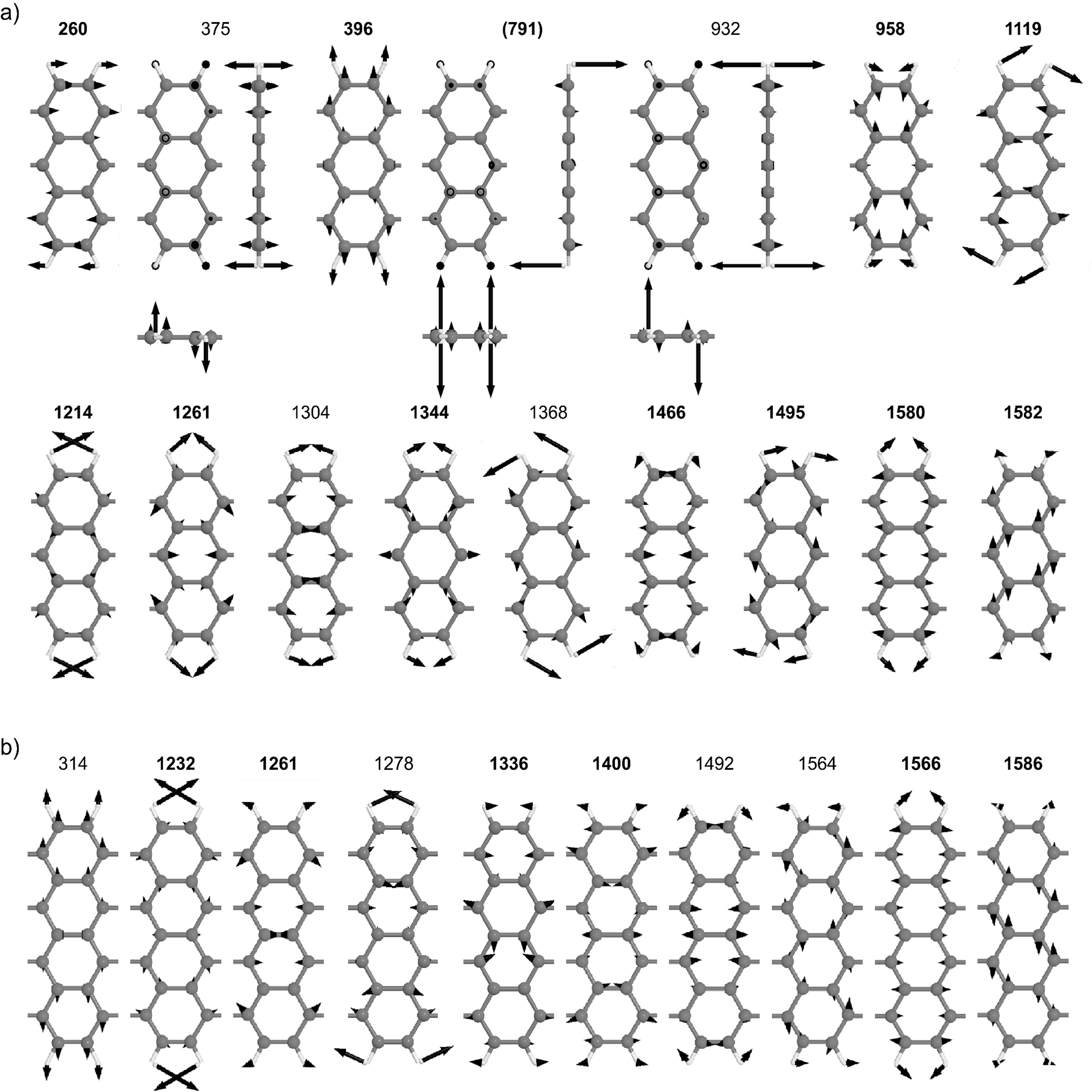}
    \caption{\textbf{Normal modes calculated from periodic DFT calculations.} Atomic displacements of (a) 7- and (b) 9-AGNRs. Positions are given in \invcm~and the modes observed experimentally in this study are highlighted in bold. Note that the mode predicted at 791 \invcm~in parentheses has out-of-plane displacements (B\sub{3g} symmetry). Because the Raman signals are collected in a backscattering configuration, the experimentally observed peak at 790 \invcm~is more likely to be an overtone of an RBLM (fundamental mode, A\sub{g} symmetry).}
    \label{fs2-normal}
\end{figure}
\FloatBarrier

\begin{figure}[!ht]
    \centering
    \includegraphics[width=\textwidth]{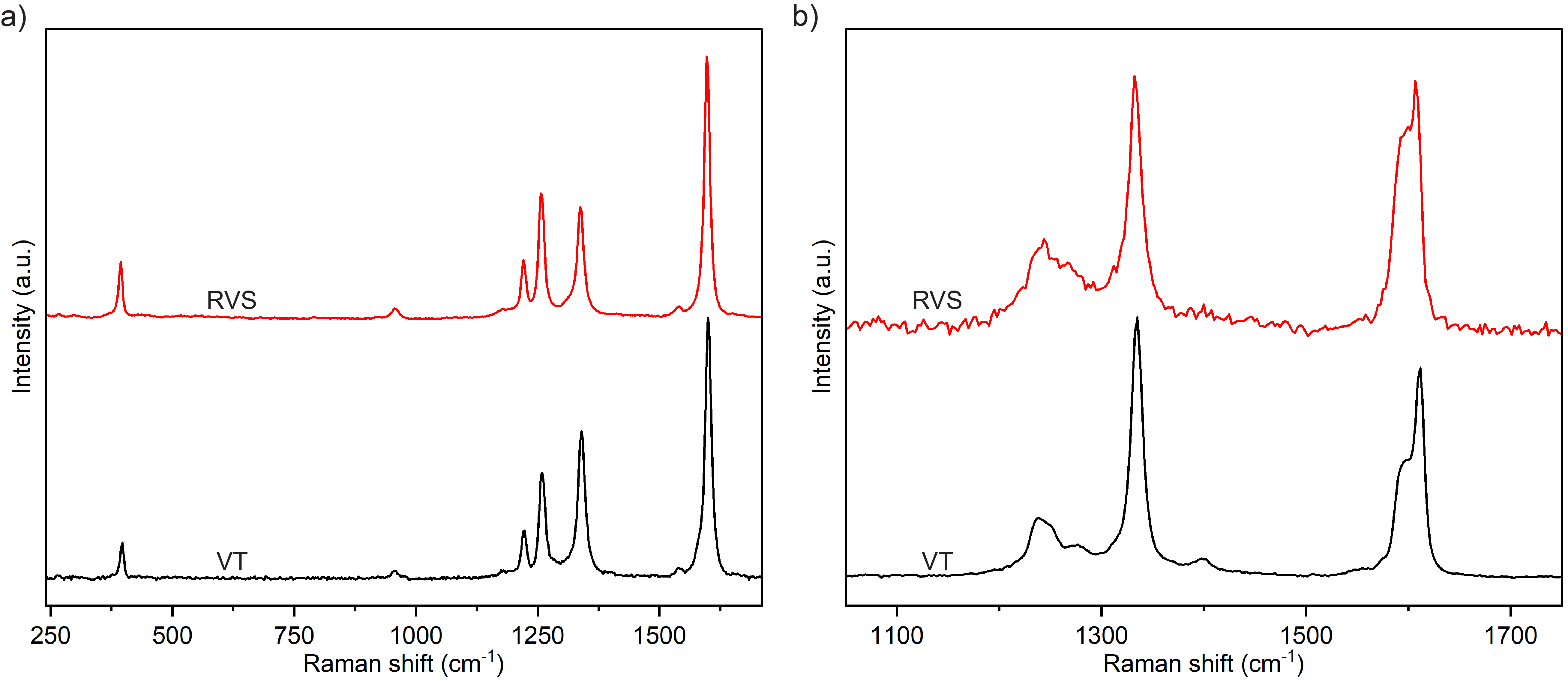}
    \caption{\textbf{Raman spectra of GNRs synthesized in the RVS and in the PC of a variable-temperature (VT) STM.} (a) 7-AGNRs and (b) 9-AGNRs. The spectra represent an average across 100 individual measurement points acquired at \excite~= 532 nm. For 7-AGNRs, the sample grown in the RVS was measured at 40 mW over a 15 \um~\mul~15 \um~area, and the sample grown in the PC of a VT STM was measured at 36 mW over a 10 \um~\mul~10 \um~area. Both were measured with a 1 s integration time. For 9-AGNRs, both samples were measured at 10 mW with a 5 s integration time, mapped over 15 \um~\mul~15 \um~(RVS) and 20 \um~\mul~20 \um~(VT).}
    \label{fs2-Raman_InSitu-vs-vt}
\end{figure}
\FloatBarrier

\begin{figure}[!ht]
    \centering
    \includegraphics[width=\textwidth]{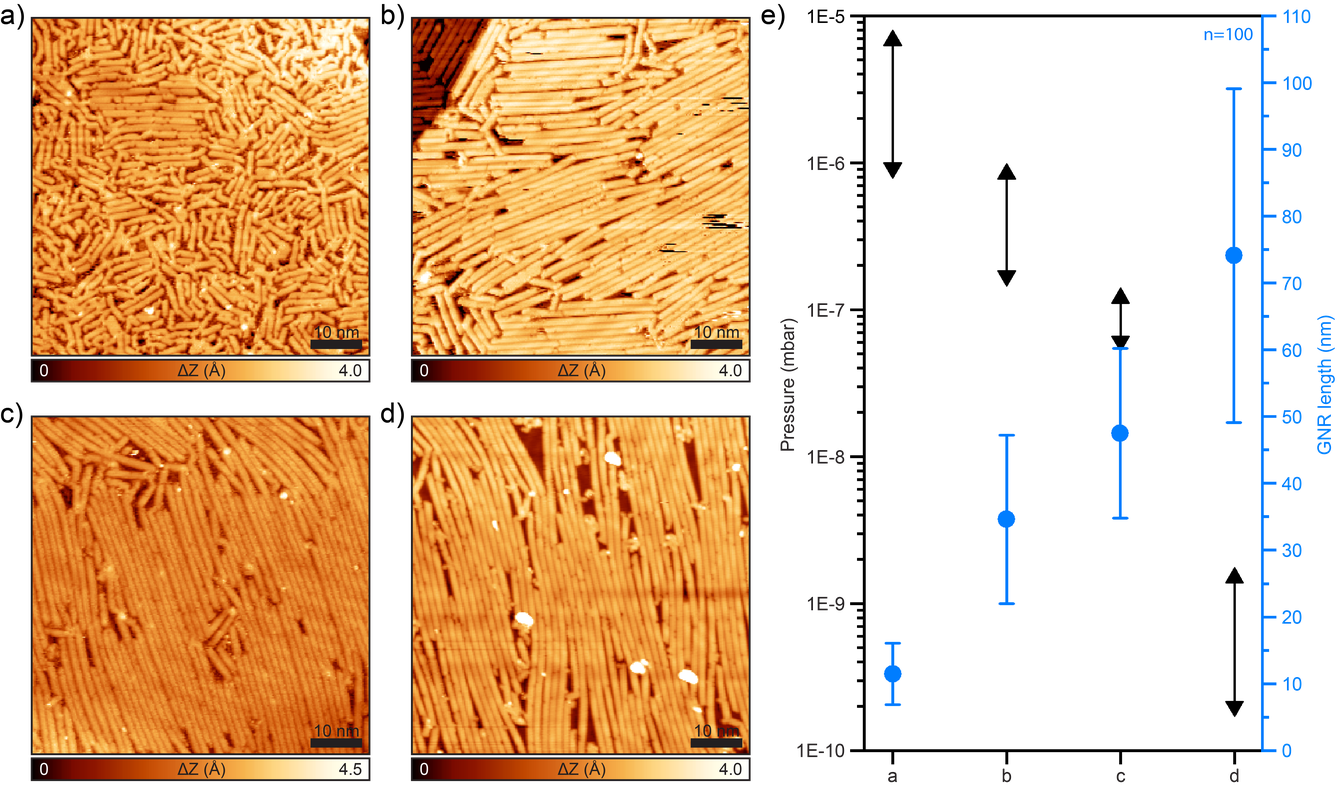}
    \caption{\textbf{Effect of chamber pressure on the length of the GNRs grown in the RVS.} STM images of 9-AGNRs grown under different conditions. (a) Gate to FEL closed, no air cooling. (b) Gate closed, with air cooling. (c) Gate open, with air cooling. (d) Reference sample grown in the PC. Scanning parameters: (a) -1 V, 25 pA, (b--d) -1 V, 30 pA. All STM images were collected at room temperature. (e) Pressure ranges during growth (black) and corresponding length distributions (blue) for each condition. Histograms of measured lengths are shown in Figure \ref{fs2-hist}.}
    \label{fs2-pressure}
\end{figure}
\FloatBarrier

\begin{figure}[!ht]
    \centering
    \includegraphics[width=\textwidth]{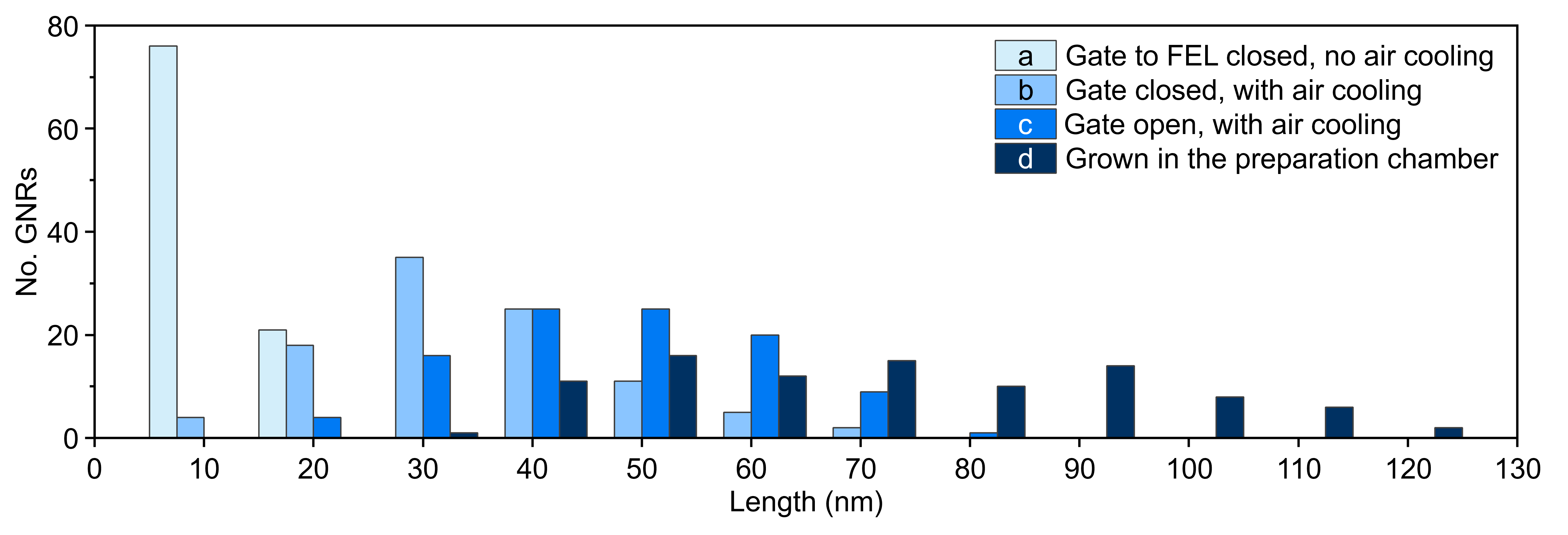}
    \caption{\textbf{Length distributions of 9-AGNRs} grown under conditions (a)--(d) in Figure \ref{fs2-pressure}. A total of 100 ribbons were measured per sample produced under each condition. Bin width: 10 nm.}
    \label{fs2-hist}
\end{figure}
\FloatBarrier

\begin{figure}[!ht]
    \centering
    \includegraphics[width=\textwidth]{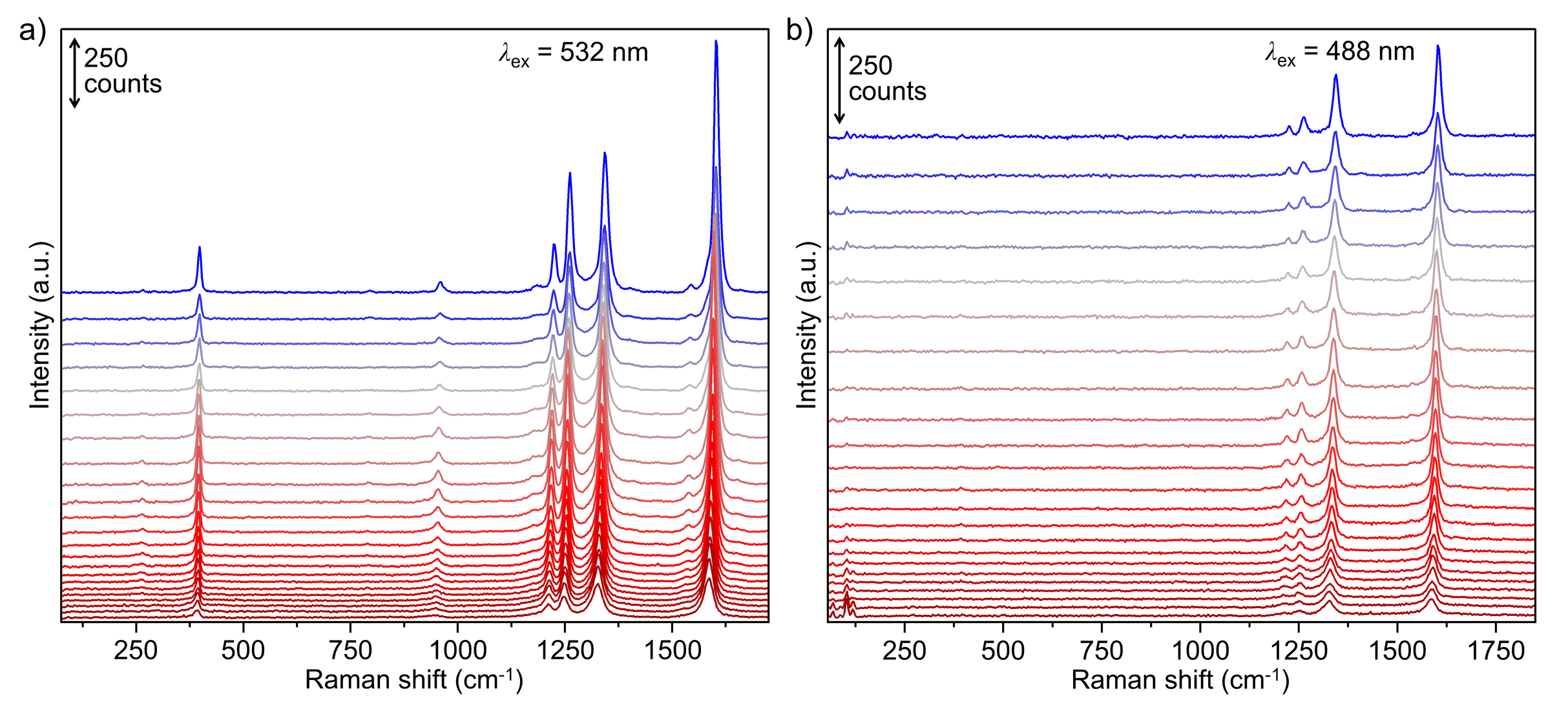}
    \caption{\textbf{Raman spectra of 7-AGNRs measured from 162 to 748 K} using (a) 532 and (b) 488 nm excitation. Raman measurement conditions: (a) \excite~= 532 nm, 40 mW, (b) \excite~= 488 nm, 10 mW. A total of 100 spectra were collected from a 15 \um~\mul~15 \um~area with a 1 s integration time and subsequently averaged. In panel (b), LCMs are observed in addition to the CH/D and G modes, whereas the RBLM is not observed.} 
    \label{fs3-1}
\end{figure}
\FloatBarrier

\begin{figure}[!ht]
    \centering
    \includegraphics[width=\textwidth]{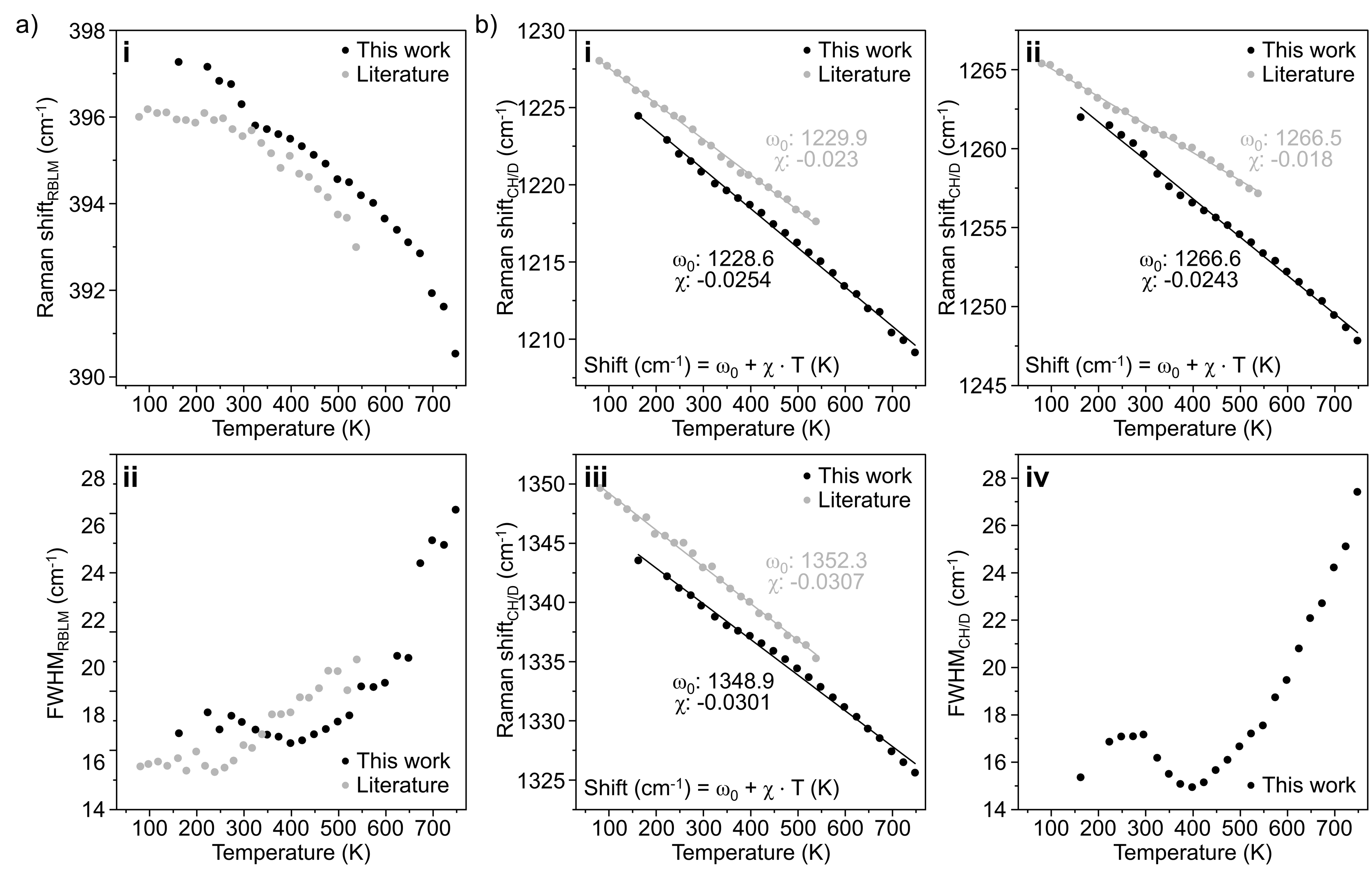}
    \caption{\textbf{Temperature dependence of Raman peak positions and linewidths of 7-AGNRs.} (a) RBLM: (i) peak position and (ii) FWHM as a function of temperature. (b) CH/D modes: (i)--(iii) peak positions and (iv) FWHM as a function of temperature. Black and gray data points correspond to values measured in this work and those reported by Guo et al.\cite{Guo2022Phonon}, respectively. See Table \ref{ts-tdep7} for complete peak information.}
    \label{fs3-trend}
\end{figure}
\FloatBarrier

\begin{figure}[!ht]
    \centering
    \includegraphics[width=0.7\textwidth]{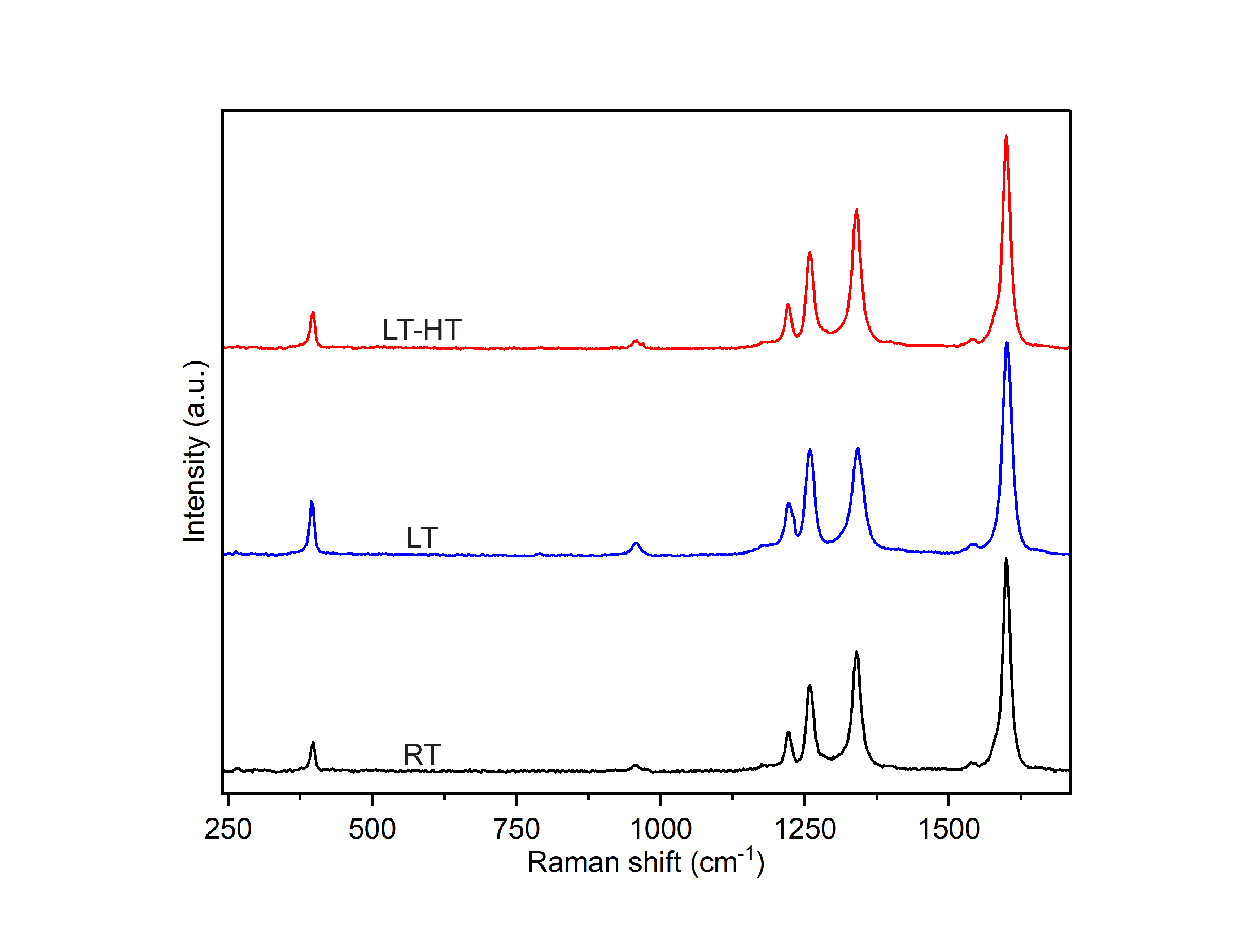}
    \caption{\textbf{Raman spectra of 7-AGNRs following exposures to low and high temperatures.} Raman spectra collected at room temperature after cooling to 162 K (blue, LT) and after subsequent heating to 748 K (red, LT-HT, thereby exposed to 162--748 K). The spectrum acquired before any thermal treatment (black, RT) is presented for comparison. All spectra were measured in the RVS at \excite~= 532 nm, 40 mW, 1 s integration time. A total of 100 spectra were collected over a 15 \um~\mul~15 \um~area and averaged.}
    \label{fs3-after-diff-temp-trend}
\end{figure}
\FloatBarrier

\begin{figure}[!ht]
    \centering
    \includegraphics[width=0.7\textwidth]{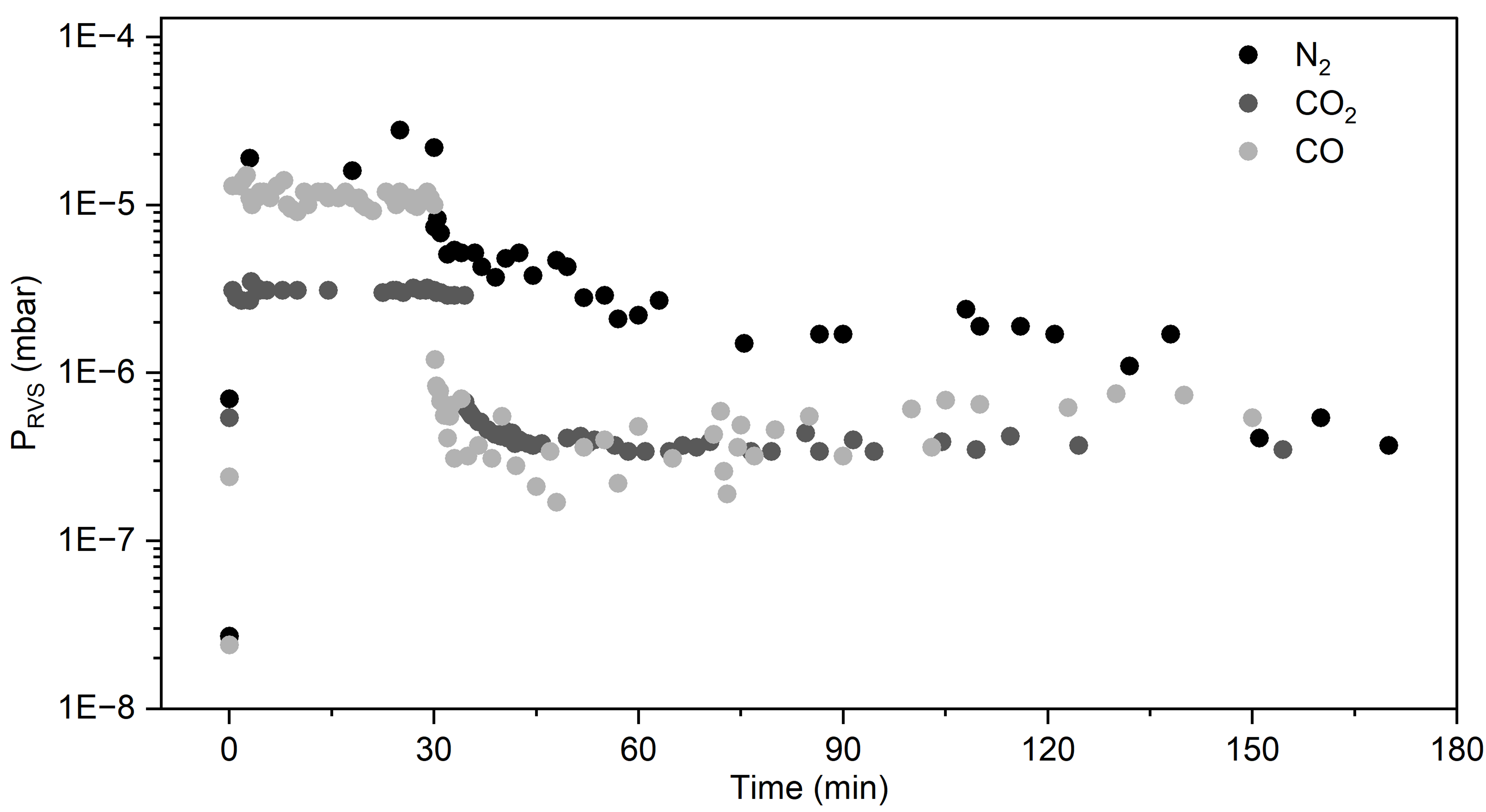}
    \caption{\textbf{Pressure recovery curves after dosing various gases in the RVS.} Starting from a base P\sub{RVS} of \ca 3 \mul~10\super{-8} mbar, \ce{N2} (black), \ce{CO2} (dark gray), and \ce{CO} (light gray) were dosed for 30 min. \ce{N2} and \ce{CO} were dosed in the 10\super{-5} mbar range and \ce{CO2} in the 10\super{-6} mbar range.}
    \label{fs4-gas}
\end{figure}
\FloatBarrier

\begin{figure}[!ht]
    \centering
    \includegraphics[width=0.8\textwidth]{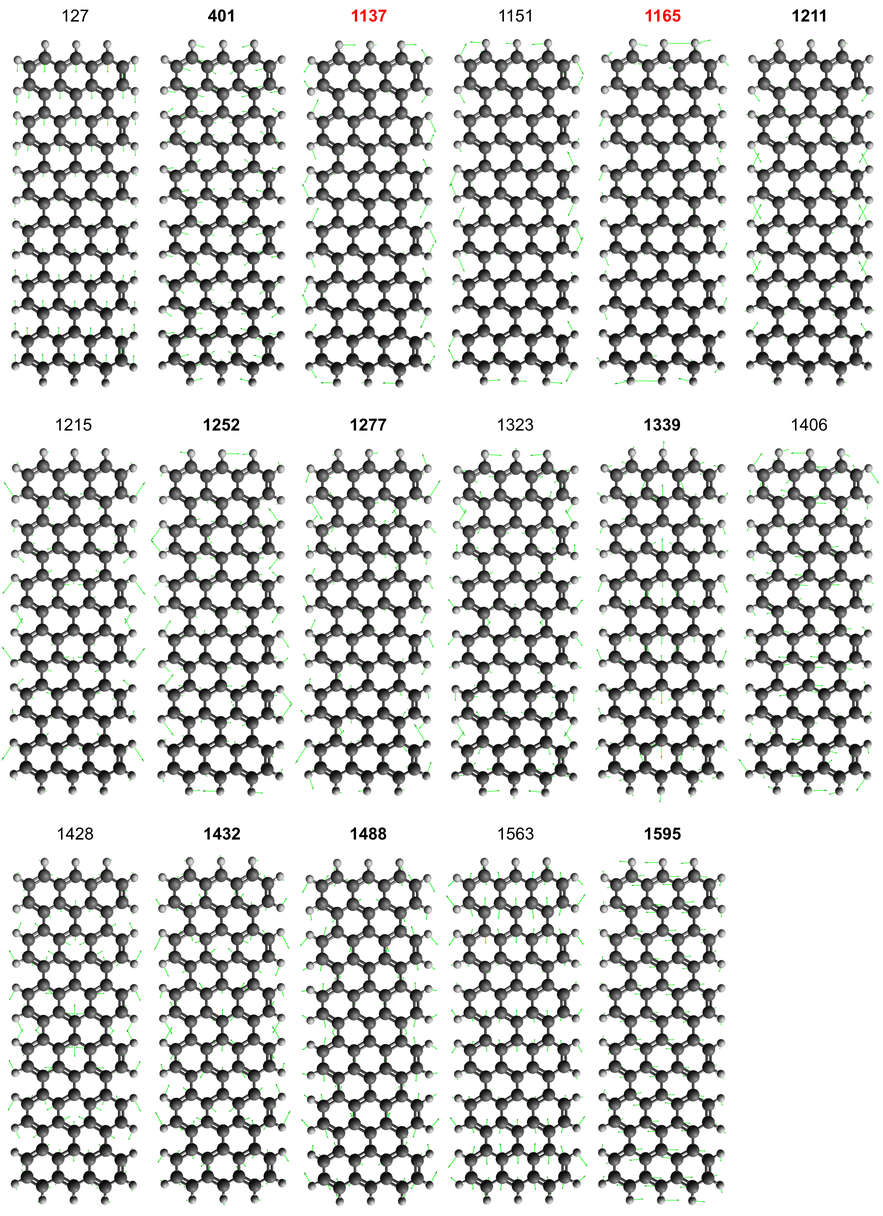}
    \caption{\textbf{Normal modes of pristine short 7-AGNRs} from a finite Raman calculation using DFT. The modes marked in Figure \ref{fs4-monobr_dft} panel i are highlighted in bold, and the modes involving large atomic displacements at the zigzag termini in red.}
    \label{fs4-s7_pr_norm}
\end{figure}
\FloatBarrier

\begin{figure}[!ht]
    \centering
    \includegraphics[width=0.7\textwidth]{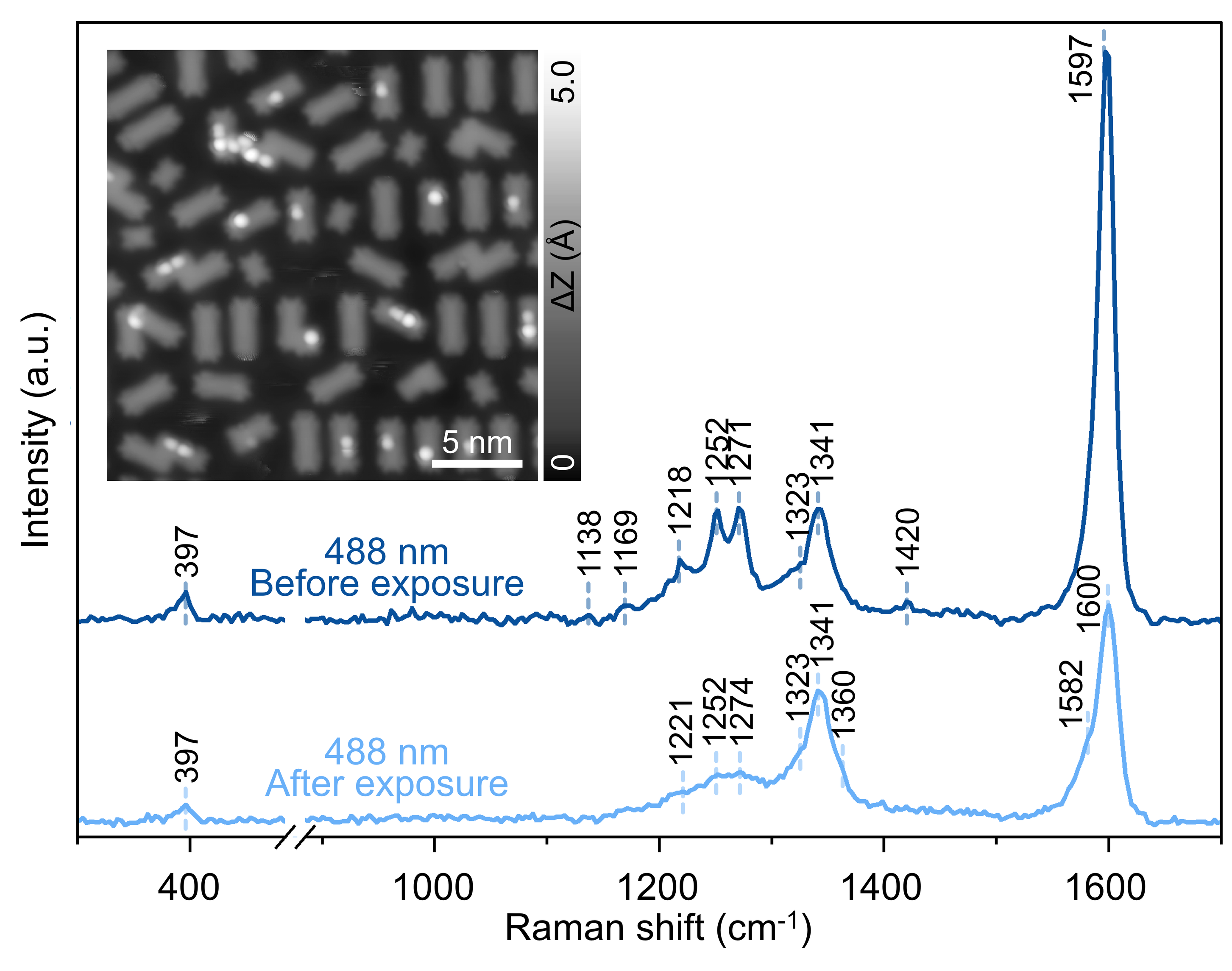}
    \caption{\textbf{Experimental Raman spectra of short 7-AGNRs before and after \ce{O2} exposure.} Raman spectra in dark and light blue correspond to those collected before and after the oxygen exposure. Measurement conditions: the spectra were acquired at 100 points over a 15 \um~\mul~15 \um~area (\excite~= 488 nm, 10 mW, 5 s integration time). 
    Inset: STM image of short 7-AGNRs (scanning parameters: -100 mV, 30 pA, 4.4 K).}
    \label{fs4-monofull}
\end{figure}
\FloatBarrier

\begin{figure}[!ht]
    \centering
    \includegraphics[width=0.7\textwidth]{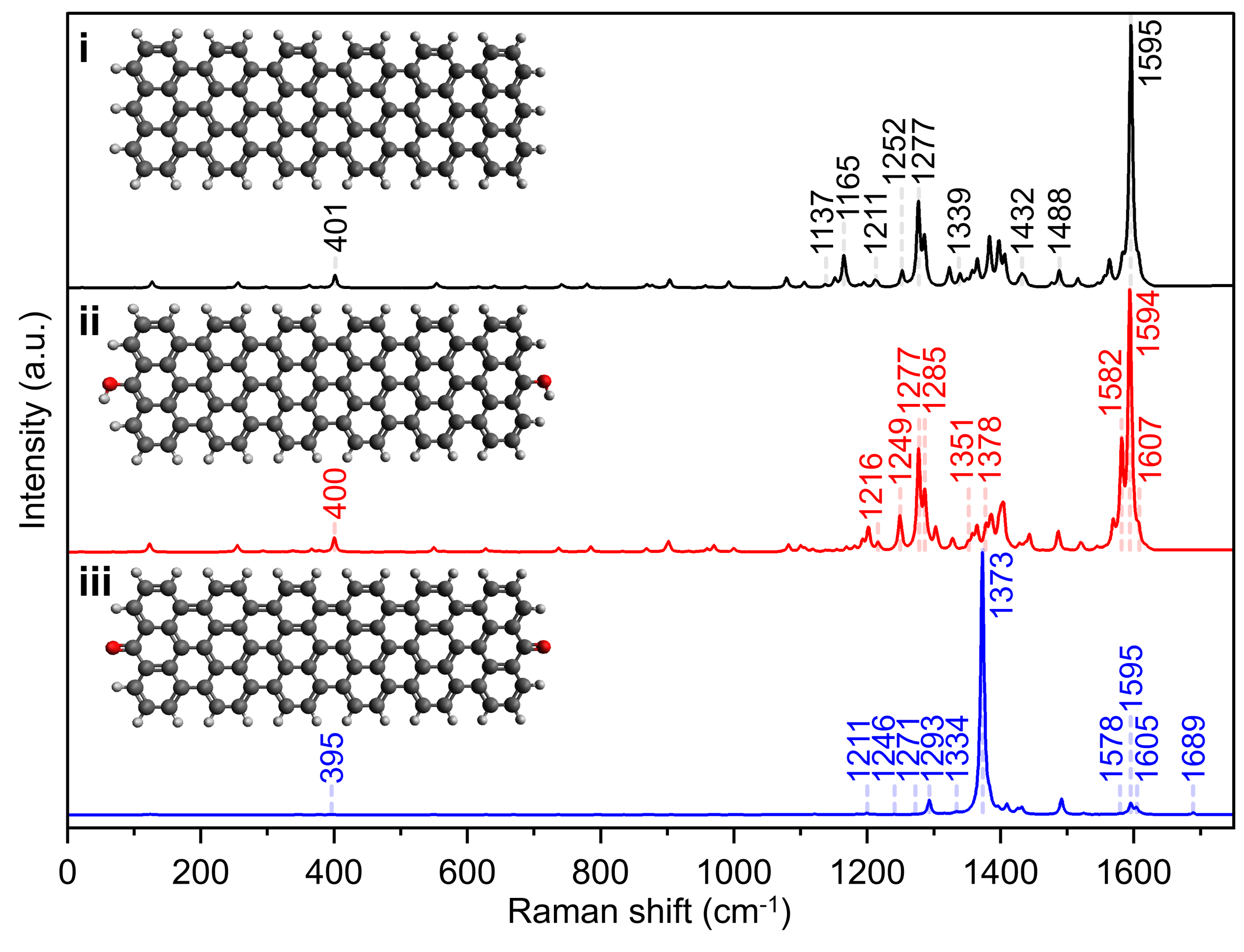}
    \caption{\textbf{Calculated Raman spectra of short 7-AGNRs} from a finite Raman calculation using DFT for (i) pristine (black), (ii) hydroxyl- (red), and (iii) ketone-substituted (blue) structures.}
    \label{fs4-monobr_dft}
\end{figure}
\FloatBarrier

\begin{figure}[!ht]
    \centering
    \includegraphics[width=0.8\textwidth]{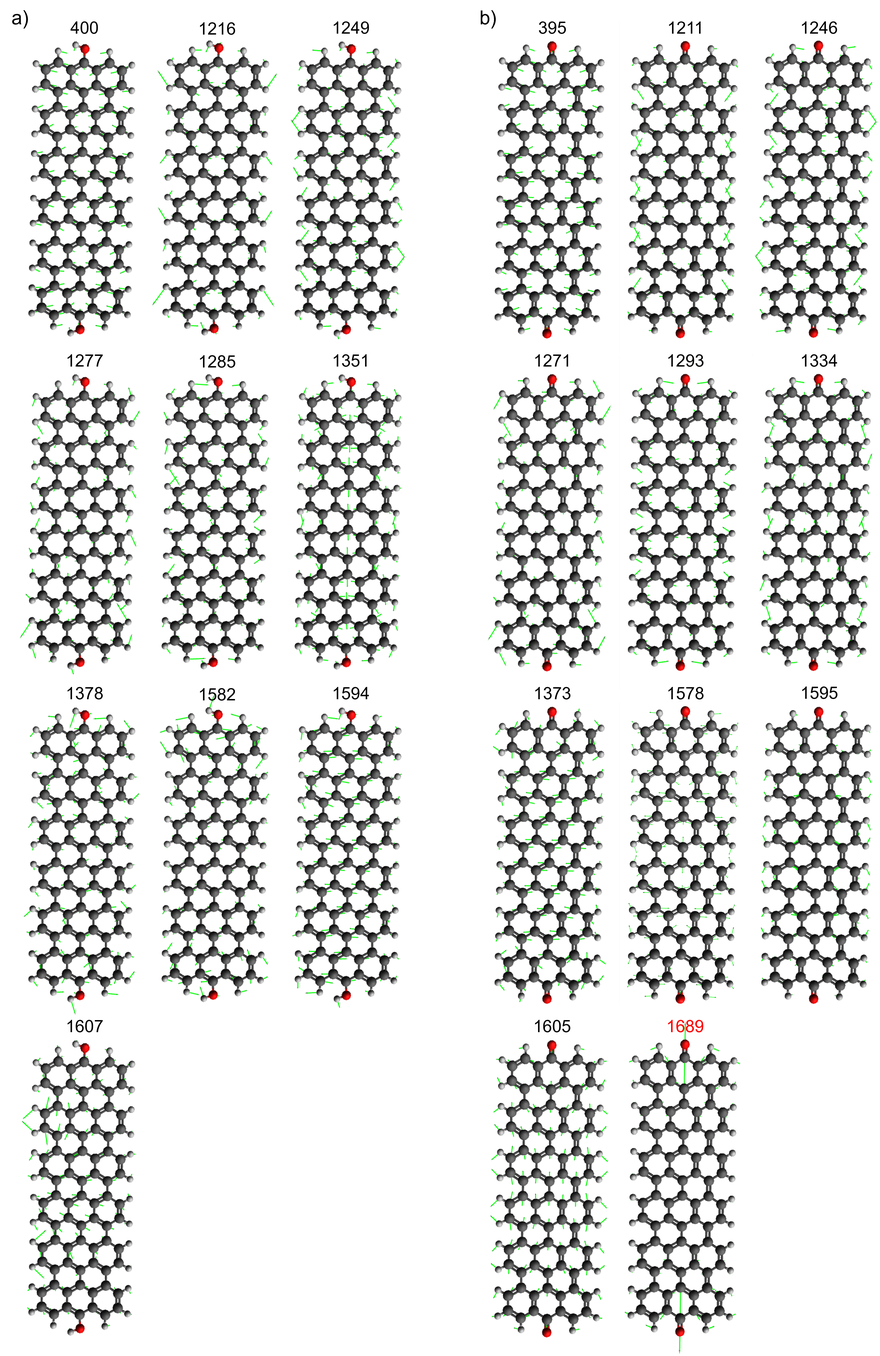}
    \caption{\textbf{Normal modes of oxidized short 7-AGNRs} functionalized with (a) \ce{-OH} and (b) \ce{=O} groups from a finite Raman calculation using DFT. In (b), the carbonyl (\ce{C=O}) stretching mode is highlighted in red. Only the modes marked in panels ii and iii of Figure \ref{fs4-monobr_dft} are displayed. Note that the \ce{C=O} stretching mode calculated at 1689 \invcm~is not observed experimentally on Au(111). See discussion below.}
    \label{fs4-s7_o2_norm}
\end{figure}
\FloatBarrier

\paragraph{DFT Raman calculations of short 7-AGNRs}
The calculated Raman spectra (Figure \ref{fs4-monobr_dft}) and the corresponding atomic displacements (Figures \ref{fs4-s7_pr_norm} and \ref{fs4-s7_o2_norm}) support the experimentally observed spectral changes. 
In the pristine short 7-AGNRs, the two modes calculated at 1137 and 1165 \invcm~involve large atomic displacements at the hydrogen atoms bound to the side and center carbons of the zigzag termini, respectively. 
Both modes are absent from the calculated spectra of the hydroxyl- and ketone-substituted structures, capturing their disappearance in the experimental spectra after exposure (Figure \ref{fs4-monofull}). 

The majority of the CH/D and G modes are retained in the calculated spectra of both substituted structures but with frequency shifts. 
The ketone-substituted structure additionally shows a mode calculated at 1689 \invcm~with carbonyl (\ce{C=O}) stretching character that is absent from both the pristine and the hydroxyl-substituted spectra. 
However, owing to the interaction between the \ce{C=O} groups and the metal, this mode is expected to become observable only when the ribbons are transferred away from the metal surface. 
In addition, short 7-AGNRs functionalized with \ce{-OH} and \ce{=O} groups are predicted to have modes at 1378 and 1582 \invcm~(\ce{-OH}) and at 1373 and 1578 \invcm~(\ce{=O}). 
The weak shoulders experimentally observed at 1360 and 1582 \invcm~can be assigned to these modes.

We note that the calculated peak intensities in Figure \ref{fs4-monobr_dft} cannot be directly compared with the experimental intensities in Figures \ref{f4}b and \ref{fs4-monofull} since the DFT calculations model isolated gas-phase molecules whereas the Raman measurements were performed on Au(111). 
Coupling of the material to the substrate\cite{Kolesov2017Low,Guo2022Substrate} as well as resonance effects associated with the excitation wavelength\cite{Sheremetyeva2024Resonant} are known to modify GNR Raman intensities and are not captured in the present calculations. 
Hence, we restrict the comparison between calculation and experiment to the presence and positions of specific modes rather than their intensities.

\begin{figure}[!ht]
    \centering
    \includegraphics[width=\textwidth]{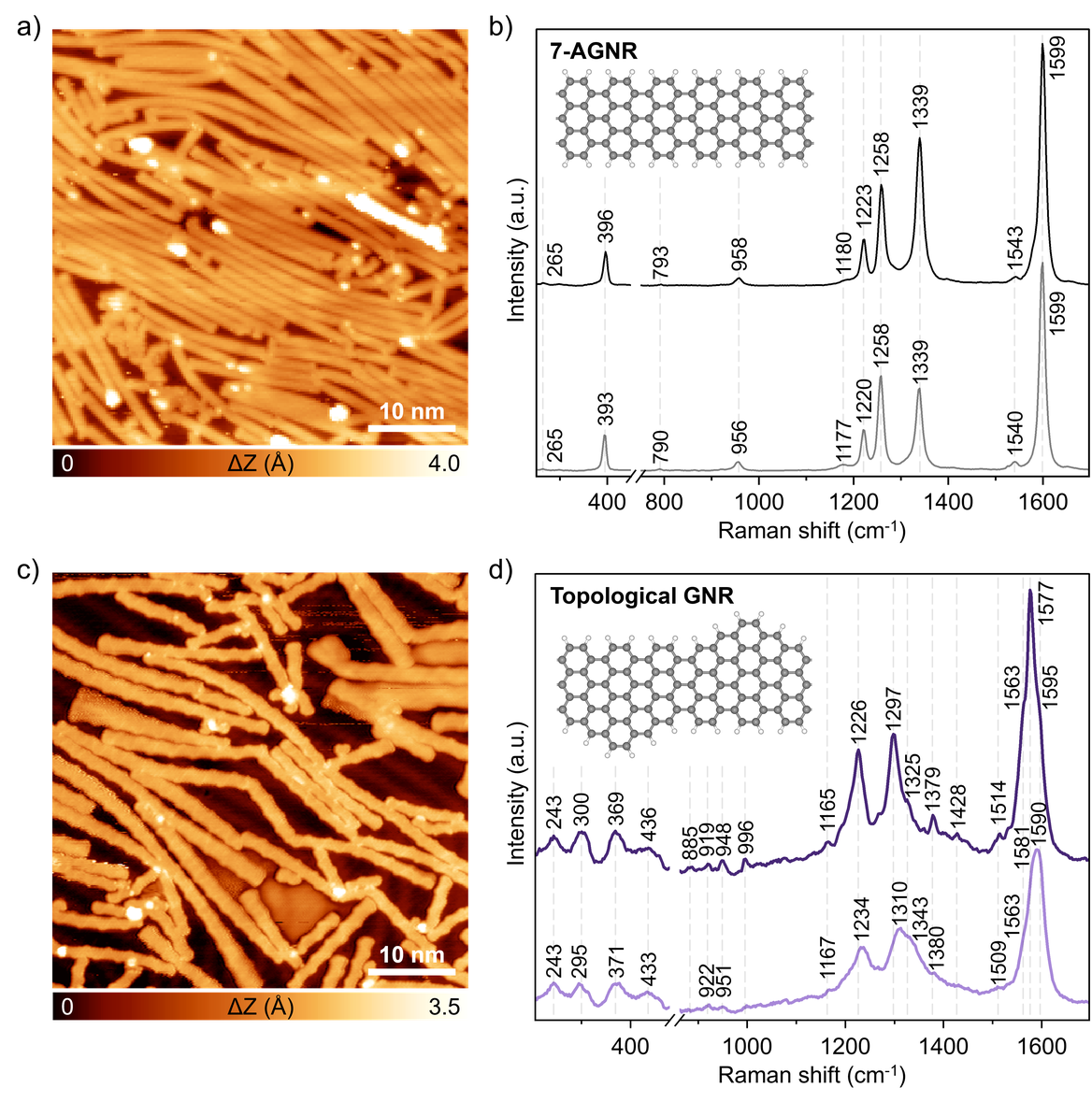}
    \caption{\textbf{STM images and Raman spectra of the 7-AGNRs and topological GNRs.} (a) STM image of the 7-AGNR sample prepared for the \ce{O2} exposure. Scanning parameters: -1 V, 20 pA, room temperature. (b) Raman spectra of 7-AGNRs before (black) and after (gray) exposure to \ce{O2}. All spectra were acquired over a 10 \um~\mul~10 \um~area (100 points, 532 nm, 35 mW, 1 s integration time). (c) STM image of the topological GNR sample. Scanning parameters: -1 V, 30 pA, room temperature. 
    (d) Raman spectra of topological GNRs before (purple) and after (light purple) exposure to \ce{O2}. All spectra were acquired over a 30 \um~\mul~30 \um~area (400 points, 532 nm, 20 mW, 5 s integration time).
    } 
    \label{fs4-topoO2}
\end{figure}
\FloatBarrier

\paragraph{Oxygen exposure of topological and 7-AGNRs}
The staggered 7-AGNR-S(1,3) (topological GNR) consists of alternating 7- and 9-AGNR segments. Topological interface states hybridize into in-gap bands, effectively reducing the bandgap from \ca 2.3 eV for pristine 7-AGNRs\cite{Ruffieux2012Electronic} to approximately 0.65 eV\cite{Groning2018Engineering}. 
A reduced bandgap is expected to lower the barrier for charge transfer to adsorbates and thereby to increase chemical reactivity. 
 
Figure \ref{fs4-topoO2}d shows the Raman spectra of topological GNRs before (purple) and after (light purple) \ce{O2} exposure. 
The two RBLM contributions from the 7-AGNR backbone and the 9-AGNR segment of the extended edges\cite{Groning2018Engineering}, centered at 369 and 300~\invcm, respectively, decrease slightly in intensity but remain largely unchanged. 
This suggests that the ribbon width is preserved.

Significant changes are nonetheless observed at higher frequencies. 
The additional breathing-like modes at 885 and 996 \invcm~disappear, along with the CH/D mode at 1428 \invcm~that arises from \ce{C-C} breathing and \ce{C-H} bending. 
The CH/D modes at 1226 and 1297 \invcm~that have a pronounced contribution from \ce{C-H} displacement, as well as the mode at 1325 \invcm, to which both \ce{C-C} and \ce{C-H} vibrations contribute, shift to 1234, 1310, and 1343 \invcm, respectively. 
Lastly, the G modes at 1577 and 1595 \invcm~shift to 1581 and 1590 \invcm.

We note that topological GNRs were previously reported to be stable upon immediate air exposure and after five days under ambient conditions\cite{Groning2018Engineering}. 
A likely origin of this discrepancy is the structural quality of the GNR sample. 
STM characterization of the topological GNR sample used in this study reveals shorter and more fused ribbons (Figure \ref{fs4-topoO2}c). 
Such structural imperfections may introduce non-hexagonal rings and disrupt the \sptwo~conjugation, making the ribbons more susceptible to oxidation\cite{Boukhvalov2008Chemical,Denis2013Comparative}, as local bond strain lowers the activation barrier\cite{Wang2011Strain}.

\begin{figure}[!ht]
    \centering
    \includegraphics[width=\textwidth]{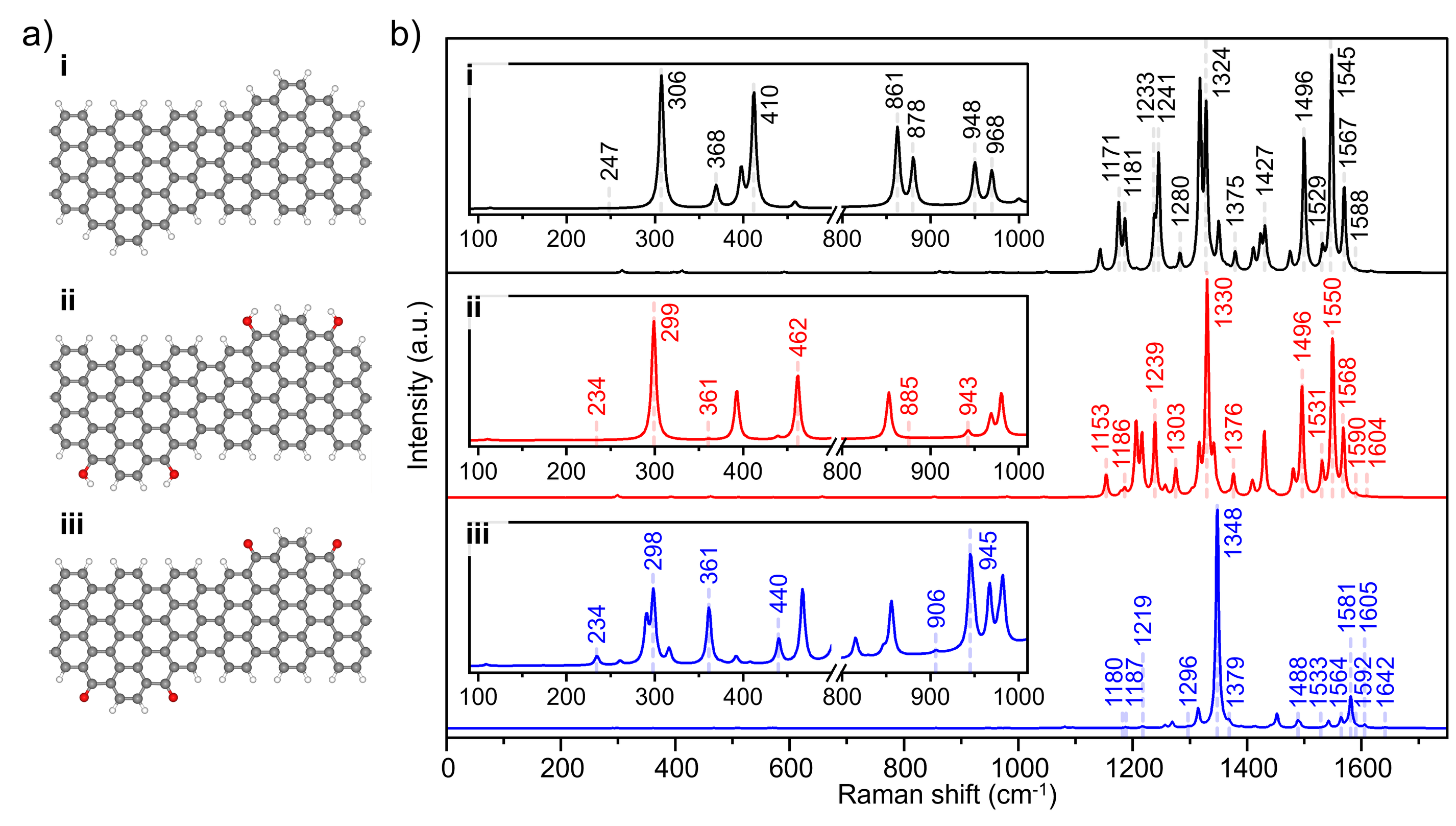}
    \caption{\textbf{Calculated Raman spectra of topological GNRs in pristine and oxidized forms.} (a) Unit cell structures of (i) pristine topological GNRs and topological GNRs functionalized with (ii) \ce{-OH} and (iii) \ce{=O} at the zigzag center of the extended edges, and (b) corresponding calculated Raman spectra.}
    \label{fs4-topo_dft}
\end{figure}
\FloatBarrier

\begin{figure}[!ht]
    \centering
    \includegraphics[width=\textwidth]{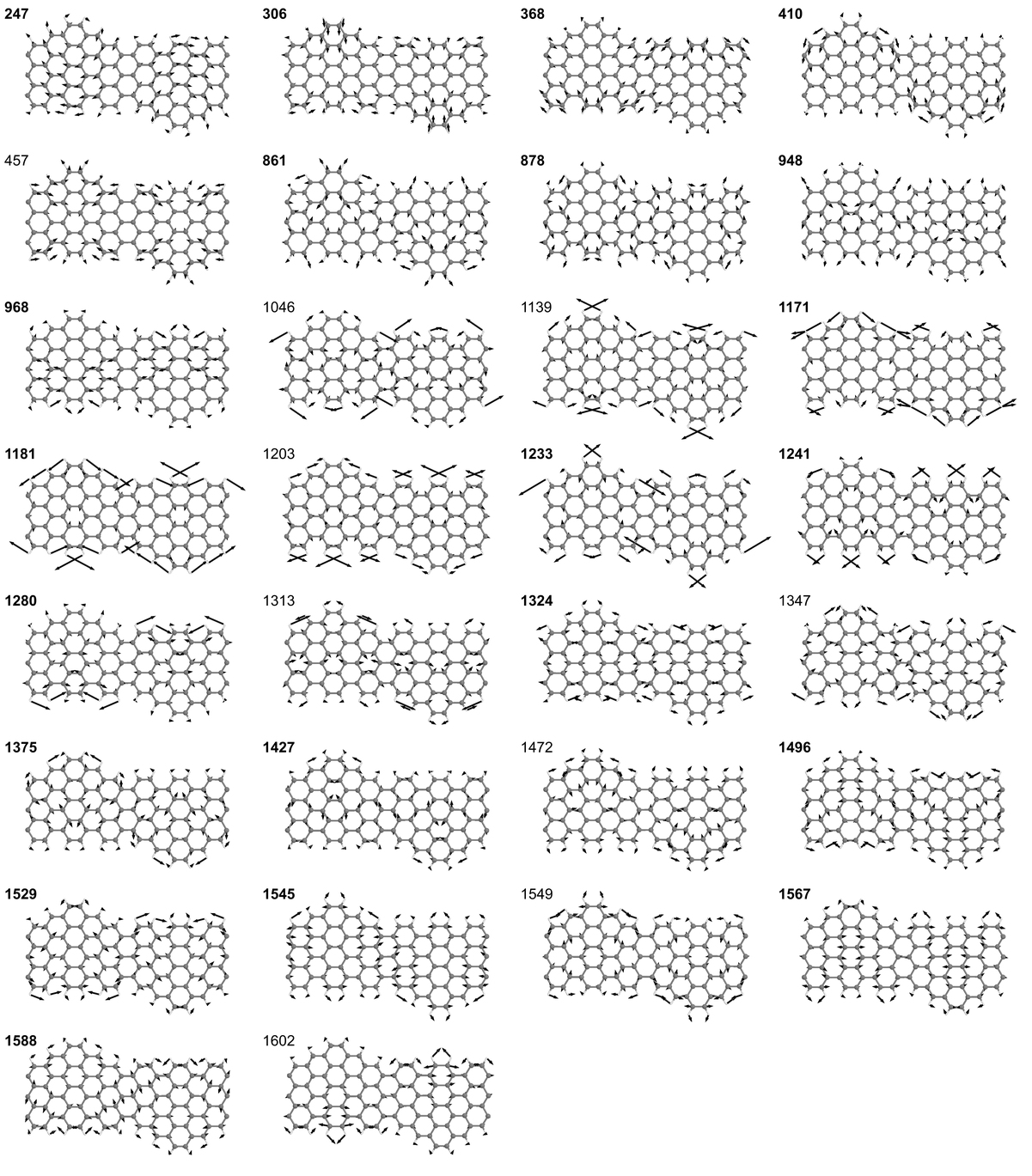}
    \caption{\textbf{Normal modes of pristine topological GNRs} from a periodic Raman calculation using DFT. The modes marked in Figure \ref{fs4-topo_dft}b panel i are highlighted in bold.}
    \label{fs4-topo_pr_norm}
\end{figure}
\FloatBarrier

\begin{figure}[!ht]
    \centering
    \includegraphics[width=0.95\textwidth]{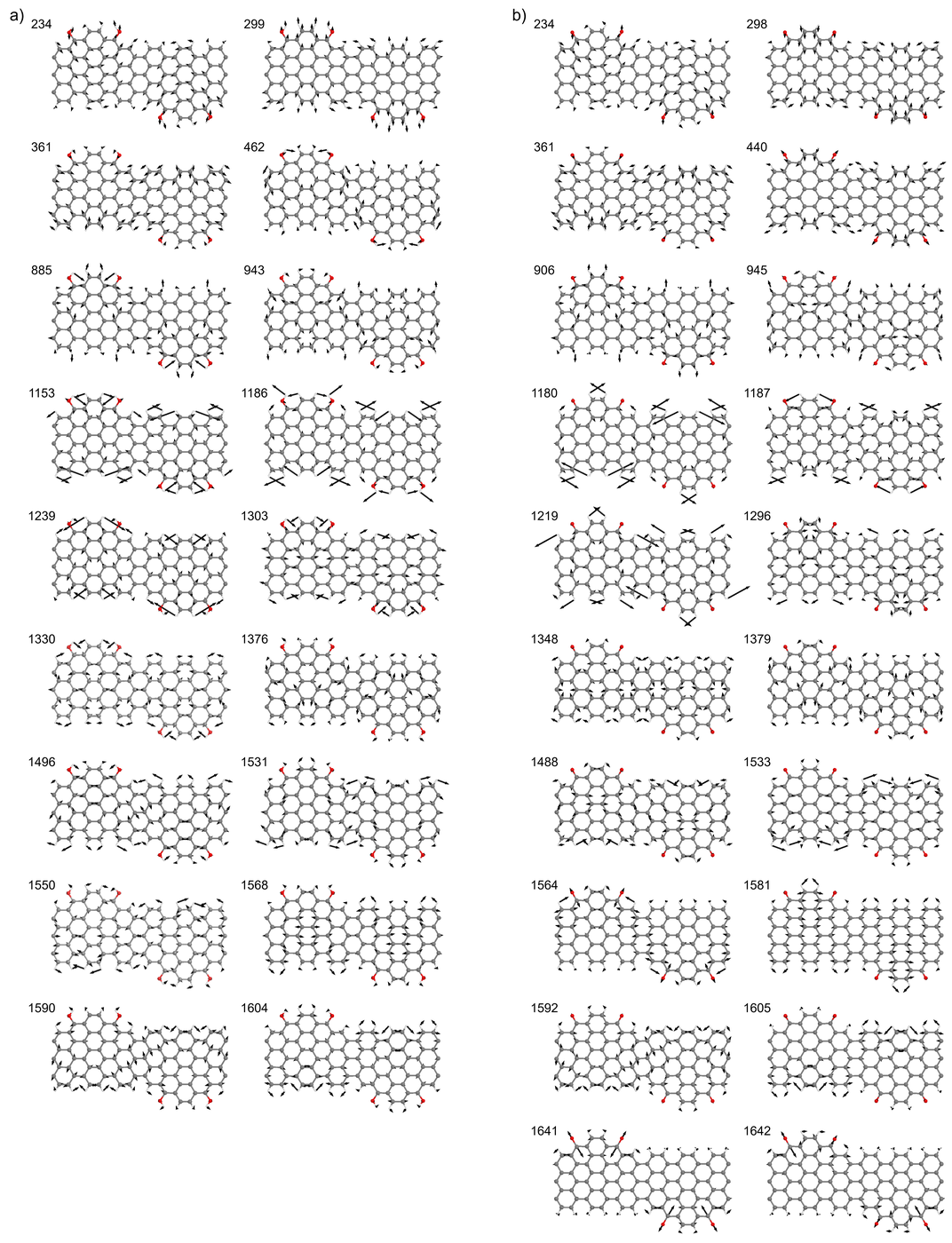}
    \caption{\textbf{Normal modes of oxidized topological GNRs} functionalized with (a) \ce{-OH} and (b) \ce{=O} groups from a periodic Raman calculation using DFT. The modes marked in Figure \ref{fs4-topo_dft}b panels ii and iii are displayed.}
    \label{fs4-topo_o2_norm}
\end{figure}
\FloatBarrier

\paragraph{DFT Raman calculations of topological GNRs}
The periodic DFT calculations (Figure \ref{fs4-topo_dft}) and the associated normal mode analysis (Figures \ref{fs4-topo_pr_norm} and \ref{fs4-topo_o2_norm}) further support the structural modification of the topological GNRs after exposure to \ce{O2}. 
\ce{-OH} and \ce{=O} groups that are introduced at the junctions between the 7- and 9-AGNR segments remove the associated \ce{C-H} bending contributions and shift the vibrational frequencies by up to a few tens of \invcm. 
Notably, the two calculated RBLMs at 306 and 368 \invcm, associated with the 9-AGNR segment and the 7-AGNR backbone, respectively, are essentially retained upon functionalization, while several of the calculated modes above 850 \invcm~are more strongly shifted. 
This is consistent with the experimentally observed persistence of the RBLM contributions alongside the more pronounced changes of the modes appearing at higher frequencies.

\clearpage

\begin{table}[!ht]
\centering
\begin{adjustbox}{max width=0.8\textwidth}
\begin{tabular}{ccrrrrrrrrrrr}
\hline
\TTC~(\degC) & DFT & 260 & 396 & (791) & 958 & 1119 & 1214 & 1261 & 1344 & 1466 & 1580 & 1582 \\
\hline
\multirow{4}{*}{248}
 & Position & & 393 & & & & 1216 & 1254 & 1333 & & & 1590 \\
 & FWHM   & & 14  & & & & 14   & 14   & 14   & & & 14   \\
 & Height & & 32  & & & & 29   & 67   & 70   & & & 157  \\
 & Area   & & 690 & & & & 650  & 1484 & 1547 & & & 3390 \\
\hline
\multirow{4}{*}{276}
 & Position & & 392 & & 957 & & 1215 & 1254 & 1332 & & & 1590 \\
 & FWHM   & & 18  & & 20  & & 17   & 17   & 17   & & & 16   \\
 & Height & & 66  & & 18  & & 83   & 166  & 199  & & & 407  \\
 & Area   & & 1851& & 573 & & 2181 & 4367 & 5250 & & & 9765 \\
\hline
\multirow{4}{*}{300}
 & Position & & 392 & & 954 & & 1214 & 1253 & 1331 & & & 1589 \\
 & FWHM   & & 10  & & 19  & & 17   & 17   & 17   & & & 16   \\
 & Height & & 101 & & 19  & & 137  & 283  & 372  & & & 675  \\
 & Area   & & 1656& & 575 & & 3608 & 7482 & 9816 & & & 16675\\
\hline
\multirow{4}{*}{328}
 & Position & & 392 & & 952 & 1182 & 1214 & 1252 & 1330 & & & 1588 \\
 & FWHM   & & 12  & & 24  & 18   & 18   & 18   & 18   & & & 17   \\
 & Height & & 105 & & 24  & 18   & 159  & 334  & 483  & & & 774  \\
 & Area   & & 1990& & 883 & 507  & 4420 & 9311 & 13438& & & 20561\\
\hline
\multirow{4}{*}{350}
 & Position & & 391 & & 952 & 1180 & 1213 & 1252 & 1329 & & & 1587 \\
 & FWHM   & & 10  & & 22  & 19   & 19   & 19   & 19   & & & 18   \\
 & Height & & 134 & & 27  & 24   & 184  & 376  & 589  & & & 882  \\
 & Area   & & 2147& & 922 & 710  & 5338 & 10927& 17110& & & 24457\\
\hline
\multirow{4}{*}{378}
 & Position & & 391 & & 950 & 1181 & 1213 & 1251 & 1328 & & & 1586 \\
 & FWHM   & & 10  & & 20  & 19   & 19   & 19   & 19   & & & 19   \\
 & Height & & 119 & & 25  & 21   & 168  & 336  & 549  & & & 778  \\
 & Area   & & 1934& & 775 & 623  & 5109 & 10188& 16616& & & 22890\\
\hline
\multirow{4}{*}{402}
 & Position & & 391 & & 951 & 1181 & 1212 & 1251 & 1327 & & & 1585 \\
 & FWHM   & & 10  & & 24  & 21   & 21   & 21   & 21   & & & 20   \\
 & Height & & 97  & & 23  & 20   & 133  & 265  & 456  & & & 618  \\
 & Area   & & 1492& & 878 & 628  & 4260 & 8522 & 14625& & & 19188\\
\hline
\multirow{4}{*}{427}
 & Position & & 391 & & 951 & 1179 & 1211 & 1250 & 1326 & & & 1584 \\
 & FWHM   & & 11  & & 23  & 22   & 22   & 22   & 22   & & & 21   \\
 & Height & & 88  & & 19  & 19   & 120  & 244  & 433  & & & 570  \\
 & Area   & & 1481& & 664 & 654  & 4022 & 8199 & 14522& & & 18648\\
\hline
\multirow{4}{*}{451}
 & Position & & 390 & & 949 & 1177 & 1211 & 1249 & 1325 & & & 1583 \\
 & FWHM   & & 10  & & 29  & 23   & 23   & 23   & 23   & & & 23   \\
 & Height & & 76  & & 16  & 20   & 99   & 202  & 370  & & & 467  \\
 & Area   & & 1146& & 696 & 707  & 3560 & 7280 & 13334& & & 16384\\
\hline
\multirow{4}{*}{480}
 & Position & & 390 & & 948 & 1175 & 1210 & 1248 & 1323 & & & 1581 \\
 & FWHM   & & 13  & & 35  & 25   & 25   & 25   & 25   & & & 25   \\
 & Height & & 56  & & 9   & 16   & 75   & 155  & 293  & & & 362  \\
 & Area   & & 1114& & 498 & 634  & 2938 & 6059 & 11480& & & 13520\\
\hline
\multirow{4}{*}{RT}
 & Position & 268 & 394 & 790 & 958 & 1178 & 1221 & 1258 & 1338 & 1539 & 1595 & 1600 \\
 & FWHM   & 12  & 8   & 16  & 17  & 15   & 15   & 15   & 15   & 12   & 12   & 12   \\
 & Height & 27  & 479 & 16  & 90  & 68   & 388  & 1034 & 984  & 67   & 1103 & 1382 \\
 & Area   & 519 & 6300& 395 & 2399& 1528 & 8783 & 23393& 22238& 1195 & 19660& 24609\\
\hline
\end{tabular}
\end{adjustbox}
\caption{\textbf{Peak parameters extracted from the in situ Raman spectra acquired during the OSS of 7-AGNRs in the RVS}. The theoretical positions are given in the first row. Units: Positions and FWHM are in \invcm, heights in counts, and areas in counts$\,\cdot\,$\invcm. RT denotes a 7-AGNR sample measured after cooling down to room temperature following the OSS. Note that the LO and TO components become discernible only upon cooling to room temperature after growth.}
\label{ts-insitu7}
\end{table}

\begin{table}[!ht]
\centering
\begin{adjustbox}{max width=0.5\textwidth}
\begin{tabular}{ccrrrrrr}
\hline
\TTC~(\degC) & DFT & 1232 & 1261 & 1336 & 1400 & 1566 & 1586 \\
\hline
\multirow{4}{*}{327}
 & Position & 1248 & & 1327 & & & 1586 \\
 & FWHM   & 48   & & 48   & & & 27   \\
 & Height & 35   & & 25   & & & 70   \\
 & Area   & 2552 & & 1807 & & & 2910 \\
\hline
\multirow{4}{*}{351}
 & Position & 1242 & 1262 & 1327 & 1389 & 1582 & 1595 \\
 & FWHM   & 31   & 31   & 31   & 31   & 18   & 18   \\
 & Height & 46   & 21   & 54   & 4    & 82   & 67   \\
 & Area   & 2200 & 991  & 2585 & 199  & 2302 & 1881 \\
\hline
\multirow{4}{*}{376}
 & Position & 1231 & 1259 & 1325 & 1375 & 1581 & 1597 \\
 & FWHM   & 23   & 23   & 23   & 23   & 15   & 15   \\
 & Height & 42   & 24   & 90   & 7    & 70   & 99   \\
 & Area   & 1454 & 856  & 3145 & 257  & 1571 & 2235 \\
\hline
\multirow{4}{*}{400}
 & Position & 1228 & 1259 & 1324 & 1373 & 1580 & 1596 \\
 & FWHM   & 22   & 22   & 22   & 22   & 16   & 16   \\
 & Height & 35   & 23   & 90   & 7    & 58   & 91   \\
 & Area   & 1181 & 779  & 3028 & 222  & 1397 & 2213 \\
\hline
\multirow{4}{*}{426}
 & Position & 1230 & 1264 & 1324 & 1373 & 1579 & 1595 \\
 & FWHM   & 22   & 22   & 22   & 22   & 15   & 15   \\
 & Height & 38   & 19   & 85   & 7    & 57   & 89   \\
 & Area   & 1308 & 665  & 2954 & 242  & 1355 & 2110 \\
\hline
\multirow{4}{*}{453}
 & Position & 1229 & 1258 & 1323 & 1373 & 1579 & 1594 \\
 & FWHM   & 25   & 25   & 25   & 25   & 17   & 17   \\
 & Height & 31   & 20   & 83   & 8    & 59   & 83   \\
 & Area   & 1209 & 779  & 3204 & 309  & 1514 & 2142 \\
\hline
\multirow{4}{*}{476}
 & Position & 1228 & 1260 & 1322 & 1372 & 1578 & 1593 \\
 & FWHM   & 26   & 26   & 26   & 26   & 18   & 18   \\
 & Height & 31   & 19   & 81   & 6    & 56   & 77   \\
 & Area   & 1228 & 760  & 3195 & 226  & 1534 & 2102 \\
\hline
\multirow{4}{*}{RT}
 & Position & 1241 & 1270 & 1333 & 1391 & 1592 & 1607 \\
 & FWHM   & 20   & 20   & 20   & 20   & 13   & 13   \\
 & Height & 61   & 39   & 150  & 15   & 116  & 147  \\
 & Area   & 1905 & 1215 & 4699 & 480  & 2420 & 3070 \\
\hline
\end{tabular}
\end{adjustbox}
\caption{\textbf{Peak parameters extracted from the in situ Raman spectroscopy during the OSS of 9-AGNRs in the RVS.} The theoretical positions are given in the first row. Units: Positions and FWHM are in \invcm, heights in counts, and areas in counts$\,\cdot\,$\invcm. RT denotes a 9-AGNR sample measured after cooling down to room temperature following the OSS.}
\label{ts-insitu9}
\end{table}

\begin{table}[!ht]
\centering
\begin{adjustbox}{max width=0.565\textwidth}
\begin{tabular}{ccrrrrrrrrrrrr}
\hline
\TTC~(K) & DFT & 260 & 396 & (791) & 958 & 1119 & 1214 & 1261 & 1344 & 1466 & 1495 & 1580 & 1582 \\
\hline
\multirow{4}{*}{162}
 & Position & 266 & 397 & 795 & 960 & 1182 & 1224 & 1262 & 1344 & 1543 & 1583 & 1601 & 1607 \\
 & FWHM   & 13  & 9   & 15  & 14  & 15   & 15   & 15   & 15   & 10   & 10   & 10   & 10   \\
 & Height & 9   & 168 & 7   & 38  & 17   & 127  & 395  & 527  & 17   & 48  & 669  & 430  \\
 & Area   & 183 & 2247& 168 & 839 & 402  & 3034 & 9464 & 12596& 282  & 776  & 10833& 6958 \\
\hline
\multirow{4}{*}{223}
 & Position & 266 & 397 & 794 & 959 & 1182 & 1223 & 1261 & 1342 & 1542 & 1581 & 1600 & 1606 \\
 & FWHM   & 9   & 9   & 16  & 17  & 17   & 17   & 17   & 17   & 12   & 12   & 12   & 12   \\
 & Height & 6   & 90  & 8   & 20  & 9  & 72   & 222  & 342  & 8    & 33   & 376  & 252  \\
 & Area   & 93  & 1308& 189 & 524 & 248  & 1890 & 5839 & 8964 & 152  & 595  & 6837 & 4577 \\
\hline
\multirow{4}{*}{248}
 & Position & 264 & 397 & & 958 & 1182 & 1222 & 1261 & 1341 & 1541 & 1581 & 1600 & 1605 \\
 & FWHM   & 15  & 9   & & 15  & 17   & 17   & 17   & 17   & 12   & 12   & 12   & 12   \\
 & Height & 6   & 108 & & 22  & 11   & 87   & 257  & 404  & 13   & 41   & 436  & 280  \\
 & Area   & 152 & 1469& & 520 & 293  & 2305 & 6841 & 10725& 235  & 767  & 8077 & 5189 \\
\hline
\multirow{4}{*}{273}
 & Position & 267 & 397 & & 958 & 1181 & 1222 & 1260 & 1341 & 1541 & 1580 & 1599 & 1604 \\
 & FWHM   & 10  & 9   & & 15  & 17   & 17   & 17   & 17   & 12   & 12   & 12   & 12   \\
 & Height & 6   & 105 & & 21  & 11   & 85   & 243  & 384  & 10   & 37   & 342  & 289  \\
 & Area   & 96  & 1502& & 512 & 282  & 2251 & 6463 & 10221& 192  & 700  & 6472 & 5465 \\
\hline
\multirow{4}{*}{295}
 & Position & 265 & 396 & 796 & 957 & 1179 & 1221 & 1260 & 1340 & 1540 & 1579 & 1598 & 1603 \\
 & FWHM   & 14  & 9   & 15  & 16  & 17   & 17   & 17   & 17   & 12   & 12   & 12   & 12   \\
 & Height & 4   & 101 & 4   & 22  & 11   & 85   & 241  & 382  & 11   & 38   & 365  & 292  \\
 & Area   & 79  & 1418& 96  & 537 & 291  & 2268 & 6436 & 10214& 209  & 731  & 6950 & 5553 \\
\hline
\multirow{4}{*}{324}
 & Position & 264 & 396 & 792 & 956 & 1178 & 1220 & 1258 & 1339 & 1539 & 1578 & 1596 & 1601 \\
 & FWHM   & 17  & 9   & 16  & 17  & 16   & 16   & 16   & 16   & 12   & 12   & 12   & 12   \\
 & Height & 5   & 136 & 5   & 28  & 15   & 110  & 295  & 415  & 15   & 26   & 381  & 384  \\
 & Area   & 141 & 1853& 125 & 740 & 380  & 2780 & 7433 & 10451& 272  & 477  & 7109 & 7160 \\
\hline
\multirow{4}{*}{349}
 & Position & 264 & 396 & 792 & 955 & 1178 & 1220 & 1258 & 1338 & 1539 & & 1595 & 1600 \\
 & FWHM   & 9   & 9   & 11  & 16  & 16   & 16   & 16   & 16   & 12   & & 12   & 12   \\
 & Height & 8   & 172 & 5   & 38  & 21   & 144  & 362  & 458  & 21   & & 420  & 483  \\
 & Area   & 110 & 2288& 91  & 959 & 510  & 3485 & 8759 & 11061& 386  & & 7813 & 8993 \\
\hline
\multirow{4}{*}{373}
 & Position & 263 & 396 & 791 & 954 & 1177 & 1219 & 1257 & 1338 & 1539 & & 1595 & 1600 \\
 & FWHM   & 10  & 8   & 13  & 17  & 15   & 15   & 15   & 15   & 12   & & 12   & 12   \\
 & Height & 12  & 188 & 6   & 40  & 20   & 159  & 384  & 459  & 20   & & 447  & 469  \\
 & Area   & 183 & 2487& 125 & 1078& 481  & 3729 & 9023 & 10789& 360  & & 8206 & 8601 \\
\hline
\multirow{4}{*}{398}
 & Position & 263 & 395 & 792 & 954 & 1175 & 1219 & 1257 & 1337 & 1538 & & 1594 & 1599 \\
 & FWHM   & 11  & 8   & 13  & 18  & 15   & 15   & 15   & 15   & 12   & & 12   & 12   \\
 & Height & 14  & 227 & 8   & 49  & 26   & 196  & 443  & 511  & 26   & & 507  & 533  \\
 & Area   & 238 & 2916& 167 & 1345& 606  & 4558 & 10331& 11904& 461  & & 9062 & 9530 \\
\hline
\multirow{4}{*}{423}
 & Position & 264 & 395 & 790 & 953 & 1176 & 1218 & 1256 & 1337 & 1538 & & 1594 & 1599 \\
 & FWHM   & 10  & 8   & 11  & 18  & 15   & 15   & 15   & 15   & 12   & & 12   & 12   \\
 & Height & 14  & 212 & 7   & 47  & 25   & 189  & 428  & 510  & 25   & & 472  & 548  \\
 & Area   & 219 & 2765& 124 & 1310& 584  & 4473 & 10107& 12026& 460  & & 8691 & 10083\\
\hline
\multirow{4}{*}{448}
 & Position & 262 & 395 & 789 & 952 & 1176 & 1217 & 1256 & 1336 & 1538 & & 1593 & 1598 \\
 & FWHM   & 21  & 9   & 15  & 21  & 16   & 16   & 16   & 16   & 12   & & 12   & 12   \\
 & Height & 10  & 161 & 5   & 36  & 22   & 151  & 332  & 421  & 19   & & 335  & 470  \\
 & Area   & 319 & 2148& 126 & 1148& 532  & 3695 & 8105 & 10262& 364  & & 6415 & 8997 \\
\hline
\multirow{4}{*}{473}
 & Position & 263 & 395 & 789 & 952 & 1175 & 1217 & 1255 & 1335 & 1537 & & 1592 & 1597 \\
 & FWHM   & 25  & 9   & 13  & 21  & 16   & 16   & 16   & 16   & 13   & & 13   & 13   \\
 & Height & 10  & 151 & 4   & 33  & 19   & 142  & 307  & 399  & 17   & & 292  & 448  \\
 & Area   & 382 & 2052& 84  & 1101& 489  & 3567 & 7705 & 10003& 328  & & 5714 & 8774 \\
\hline
\multirow{4}{*}{498}
 & Position & 259 & 395 & 789 & 952 & 1175 & 1216 & 1255 & 1334 & 1536 & & 1591 & 1596 \\
 & FWHM   & 31  & 9   & 13  & 22  & 17   & 17   & 17   & 17   & 14   & & 14   & 14   \\
 & Height & 8   & 121 & 5   & 27  & 16   & 118  & 248  & 337  & 17   & & 220  & 366  \\
 & Area   & 383 & 1701& 103 & 920 & 411  & 3057 & 6432 & 8750 & 372  & & 4671 & 7768 \\
\hline
\multirow{4}{*}{523}
 & Position & 263 & 394 & 788 & 951 & 1175 & 1216 & 1254 & 1334 & 1535 & & 1590 & 1595 \\
 & FWHM   & 21  & 9   & 7   & 22  & 17   & 17   & 17   & 17   & 14   & & 14   & 14   \\
 & Height & 11  & 109 & 5   & 24  & 14   & 109  & 226  & 318  & 14   & & 189  & 351  \\
 & Area   & 364 & 1559& 58  & 825 & 384  & 2912 & 6051 & 8511 & 308  & & 4077 & 7553 \\
\hline
\multirow{4}{*}{548}
 & Position & 261 & 394 & 791 & 950 & 1176 & 1215 & 1253 & 1333 & 1536 & & 1590 & 1595 \\
 & FWHM   & 19  & 10  & 8   & 22  & 18   & 18   & 18   & 18   & 15   & & 15   & 15   \\
 & Height & 9   & 105 & 5   & 25  & 15   & 107  & 218  & 314  & 11   & & 221  & 263  \\
 & Area   & 259 & 1669& 57  & 827 & 410  & 2922 & 5960 & 8576 & 251  & & 4984 & 5914 \\
\hline
\multirow{4}{*}{574}
 & Position & 260 & 394 & & 949 & 1173 & 1214 & 1253 & 1332 & 1534 & & 1587 & 1594 \\
 & FWHM   & 17  & 10  & & 21  & 19   & 19   & 19   & 19   & 15   & & 15   & 15   \\
 & Height & 9   & 90  & & 22  & 12   & 96   & 192  & 293  & 12   & & 148  & 315  \\
 & Area   & 235 & 1423& & 721 & 359  & 2808 & 5592 & 8549 & 289  & & 3440 & 7318 \\
\hline
\multirow{4}{*}{598}
 & Position & 260 & 394 & & 949 & 1173 & 1213 & 1252 & 1331 & 1534 & & 1586 & 1593 \\
 & FWHM   & 17  & 10  & & 22  & 19   & 19   & 19   & 19   & 15   & & 15   & 15   \\
 & Height & 7   & 83  & & 21  & 12   & 88   & 173  & 274  & 11   & & 128  & 286  \\
 & Area   & 177 & 1335& & 720 & 373  & 2678 & 5230 & 8289 & 266  & & 3030 & 6751 \\
\hline
\multirow{4}{*}{624}
 & Position & 261 & 393 & & 947 & 1173 & 1213 & 1252 & 1330 & 1534 & & 1584 & 1592 \\
 & FWHM   & 20  & 11  & & 25  & 21   & 21   & 21   & 21   & 16   & & 16   & 16   \\
 & Height & 6   & 67  & & 15  & 12   & 75   & 145  & 240  & 10   & & 104  & 256  \\
 & Area   & 176 & 1177& & 601 & 381  & 2420 & 4685 & 7749 & 254  & & 2567 & 6320 \\
\hline
\multirow{4}{*}{648}
 & Position & 260 & 393 & & 947 & 1169 & 1212 & 1251 & 1329 & 1534 & & 1583 & 1591 \\
 & FWHM   & 27  & 11  & & 30  & 22   & 22   & 22   & 22   & 17   & & 17   & 17   \\
 & Height & 6   & 62  & & 15  & 11   & 67   & 127  & 219  & 10   & & 98   & 218  \\
 & Area   & 265 & 1073& & 709 & 378  & 2303 & 4369 & 7516 & 264  & & 2571 & 5705 \\
\hline
\multirow{4}{*}{673}
 & Position & & 393 & & 948 & 1174 & 1212 & 1250 & 1329 & 1532 & & 1582 & 1590 \\
 & FWHM   & & 14  & & 23  & 23   & 23   & 23   & 23   & 18   & & 18   & 18   \\
 & Height & & 43  & & 14  & 7    & 56   & 103  & 184  & 6    & & 83   & 175  \\
 & Area   & & 968 & & 511 & 263  & 1961 & 3644 & 6499 & 159  & & 2314 & 4864 \\
\hline
\multirow{4}{*}{698}
 & Position & & 392 & & 946 & 1171 & 1210 & 1249 & 1327 & 1532 & & 1580 & 1589 \\
 & FWHM   & & 15  & & 27  & 24   & 24   & 24   & 24   & 19   & & 19   & 19   \\
 & Height & & 46  & & 12  & 9    & 55   & 101  & 186  & 6    & & 76   & 181  \\
 & Area   & & 1075& & 493 & 352  & 2059 & 3803 & 6987 & 181  & & 2181 & 5165 \\
\hline
\multirow{4}{*}{723}
 & Position & & 392 & & 945 & 1171 & 1210 & 1249 & 1326 & 1530 & & 1578 & 1588 \\
 & FWHM   & & 15  & & 28  & 25   & 25   & 25   & 25   & 19   & & 19   & 19   \\
 & Height & & 38  & & 11  & 8    & 47   & 84   & 162  & 4    & & 59   & 160  \\
 & Area   & & 882 & & 467 & 306  & 1818 & 3276 & 6310 & 121  & & 1712 & 4666 \\
\hline
\multirow{4}{*}{748}
 & Position & & 391 & & 941 & 1169 & 1209 & 1248 & 1326 & 1531 & & 1577 & 1588 \\
 & FWHM   & & 16  & & 43  & 27   & 27   & 27   & 27   & 21   & & 21   & 21   \\
 & Height & & 26  & & 8   & 7    & 33   & 62   & 120  & 4    & & 50   & 110  \\
 & Area   & & 658 & & 546 & 313  & 1421 & 2618 & 5083 & 142  & & 1628 & 3604 \\
\hline
\multirow{4}{*}{296}
 & Position & 263 & 395 & 790 & 957 & 1178 & 1222 & 1260 & 1342 & 1541 & & 1598 & 1605 \\
 & FWHM   & 15  & 9   & 11  & 17  & 20   & 20   & 20   & 20   & 14   & & 14   & 14   \\
 & Height & 8   & 132 & 7   & 32  & 17   & 94   & 230  & 271  & 16   & & 340  & 267  \\
 & Area   & 182 & 1878& 114 & 841 & 522  & 2902 & 7125 & 8372 & 356  & & 7567 & 5930 \\
\hline
\end{tabular}
\end{adjustbox}
\caption{\textbf{Peak parameters extracted from temperature-dependent Raman spectra of 7-AGNRs.} The theoretical positions are given in the first row. The last row (296 K) was recorded after cooling back from 748 K. Units: Positions and FWHM are in \invcm, heights in counts, and areas in counts$\,\cdot\,$\invcm.}
\label{ts-tdep7}
\end{table}

\begin{table}[!ht]
\centering
\begin{adjustbox}{max width=\textwidth}
\begin{tabular}{cccrrrrrrrrrrrrr}
\hline
\multicolumn{2}{c}{\multirow{3}{*}{DFT}} & Pristine & 401 & 1137 & 1165 & 1211 & 1252 & 1277 & & 1339 & & 1432 & 1488 & & 1595 \\
 & & \ce{-OH}    & 400 & & & 1216 & 1249 & 1277 & 1285 & 1351 & 1378 & & & 1582 & 1594 \\
 & & \ce{=O}     & 395 & & & 1211 & 1246 & 1271 & 1293 & 1334 & 1373 & & & 1578 & 1595 \\
\hline
\multirow{4}{*}{Exp.}
 & \multirow{2}{*}{\excite~= 488 nm}
   & Pristine    & 397 & 1138 & 1169 & 1218 & 1252 & 1271 & 1323 & 1341 & & 1420 & & & 1597 \\
 & & $+$ \ce{O2} & 397 & & & 1221 & 1252 & 1274 & 1323 & 1341 & 1360 & & & 1582 & 1600 \\
 \cline{2-16}
 & \multirow{2}{*}{\excite~= 532 nm}
   & Pristine    & 399 & & 1177 & 1223 & 1253 & 1274 & & 1347 & & 1434 & 1467 & & 1599 \\
 & & $+$ \ce{O2} & & & & 1223 & 1251 & 1274 & & 1344 & & & & 1577 & 1599 \\
\hline
\end{tabular}
\end{adjustbox}
\caption{\textbf{Positions of the Raman modes from short 7-AGNRs before and after \ce{O2} exposure.} DFT values are given for the pristine ribbon, as well as after hydroxyl (\ce{-OH}) and ketone (\ce{=O}) functionalization. Experimental positions are given for \excite~= 488 nm and 532 nm excitation before and after \ce{O2} exposure. The lack of an observable RBLM at 532 nm excitation after \ce{O2} exposure can be attributed to changes in resonance conditions resulting from structural modifications. The experimental and calculated positions are in \invcm.}
\label{ts-o2-s7}
\end{table}
\FloatBarrier

\begin{sidewaystable}[!ht]
\centering
\begin{adjustbox}{max width=\textwidth}
\begin{tabular}{ccrrrrrrrrrrrrrrrrrrrrrr}
\hline
\multirow{3}{*}{DFT}
 & Pristine & 247 & 306 & 368 & 410 & 861 & 878 & 948 & 968 & 1171 & 1181 & 1233 & 1241 & 1280 & 1324 & 1375 & 1427 & 1496 & 1529 & 1545 & 1567 & 1588 & 1602 \\
 & \ce{-OH}     & 234 & 299 & 361 & 462 &     & 885 & 943 &     & 1153 & 1186 & 1239 &      & 1303 & 1330 & 1376 &      & 1496 & 1531 & 1550 & 1568 & 1590 & 1604 \\
 & \ce{=O}       & 234 & 298 & 361 & 440 &     & 906 & 945 &     & 1180 & 1187 & 1219 &      & 1296 & 1348 & 1379 &      & 1488 & 1533 & 1564 & 1581 & 1592 & 1605 \\
\hline
\multirow{2}{*}{Exp.}
 & Pristine  & 243 & 300 & 369 & 436 & 885 & 919 & 948 & 996 & 1165 & 1190 & 1226 & 1267 & 1297 & 1325 & 1379 & 1428 & 1514 & 1532 & 1563 & 1577 & 1595 & \\
 & $+$ \ce{O2} & 243 & 295 & 371 & 433 &     & 922 & 951 &     & 1167 & 1193 & 1234 &      & 1310 & 1343 & 1380 &      & 1509 & 1532 & 1563 & 1581 & 1590 & \\
\hline
\end{tabular}
\end{adjustbox}
\caption{\textbf{Positions of the Raman modes from topological GNRs before and after \ce{O2} exposure.} DFT values are given for the pristine ribbon, as well as after hydroxyl (\ce{-OH}) and ketone (\ce{=O}) functionalization. Experimental positions are given for \excite~= 532 nm excitation before and after \ce{O2} exposure. The experimental and calculated positions are in \invcm.}
\label{ts-o2-topo}
\end{sidewaystable}

\end{document}